%Paper: hep-th/9407057
%From: Robert Becker <robertb@BOURBAKI.MIT.EDU>
%Date: Mon, 11 Jul 1994 15:05:18 -0400 (EDT)

%%  This paper was typeset using plain TeX, although the macro
%%  set is somewhat unusual.  Email robertb@math.mit.edu with questions.
%%  These files come from the American Mathematical Society, ftp
%%  e-math.ams.org, directory /ams.  You need the AMS Fonts package, Version
%%  2.1 installed.  Get the amsfonts.sources.tar.Z package from the /ams
%%  directory or get the files individually in the /ams/amsfonts directory
%%  structure.  Contact your system administrator or software vendor to
%%  install these fonts if you are not comfortable doing so.
\input amssym.def
\input amssym.tex
\def\lefteqn#1{\hbox to 0pt{$\displaystyle #1$\hss}}
\def\({$}% Begin math mode in a paragraph
\def\){$}% End math mode in a paragraph
\def\[{$$}
\def\]{$$}

\parskip=1ex plus 2pt minus 1pt
\baselineskip=17pt plus 2pt minus 1pt

\font\largest=cmr10 scaled\magstep2

\def\squaretwo{\kern1pt\vbox{%
  \hrule height.3pt\hbox{\vrule width .3pt\hskip.75ex
  \vbox{\vskip1.5ex}\hskip.80ex\vrule width.3pt}
  \hrule height.3pt}\kern1pt}

\def\Cal#1{{\cal #1}}
\def\SLambda{S_\Lambda}

\def\CC{\Bbb C}
\def\gg{\frak{g}}
\def\hh{\frak{h}}
\def\hathh{\hat\hh}

\def\nn{\frak n}
\def\NN{\Bbb N}
\def\RR{\Bbb R}
\def\SS{\Bbb S}
\def\ZZ{\Bbb Z}

\def\hatrho{\hat\rho}

\def\atp{\mathop{\rm atp}\nolimits}
\def\Aut{\mathop{\rm Aut}\nolimits}

\def\ch{\mathop{\rm ch}\nolimits}
\def\diag{\mathop{\rm diag}\nolimits}
\def\M{{\rm M}}
\def\mod{\mathop{\rm mod}\nolimits}
\def\mydef{\mathop{\rm def}\nolimits}
\def\osp{\mathop{\rm osp}\nolimits}
\def\resp{\mathop{\rm resp}\nolimits}
\def\R{{\rm R}}
\def\Re{\mathop{\rm Re}\nolimits}
\def\sch{\mathop{\rm sch}\nolimits}
\def\sdim{\mathop{\rm sdim}\nolimits}
\def\sign{\mathop{\rm sign}\nolimits}
\def\Sum{\sum\limits}
\def\veca{\vec{a}}
\def\vecb{\vec{b}}
\def\veck{\vec{k}}
\def\vecn{\vec{n}}
\def\vecx{\vec{x}}

\catcode`\@=11
\def\Llleftarrow{
        \mathop{
                \lower4\p@\vbox{
                                \baselineskip4\p@ \lineskiplimit\z@
                                \kern4\p@
                                \hbox{$\leftarrow$}
                                \hbox{$\leftarrow$}
                                \hbox{$\leftarrow$}
                               }
               }
       \limits
}
\catcode`\@=\active

\def\righttypearrow#1#2{\smash{\mathop {\longrightarrow}
     \limits^{#1} \limits_{ {\raise 1ex \hbox {$\scriptstyle#2$} }
}  }}

\def\relbuilder#1#2{\smash {\mathop {\raise .5ex \hbox{$#1$}}
          \limits_{ {\raise 2ex \hbox {$\scriptstyle#2$} } }  }}

\def\blackbox{\quad\hbox{\vrule height6pt width6pt}}

\outer\def\beginsection#1#2 \par{\vskip0pt plus .2\vsize\penalty-250
     \vskip0pt plus-.2\vsize\bigskip\vskip\parskip
                          %\message{#1 #2}
                             \indent\indent\S {\bf #1.\enspace{#2.}}
                          \sectnum = #1
                          \num = 0
                          \nobreak\smallskip\indent}

\outer\def\subsection#1{\medbreak\noindent {\bf #1} \enspace}

\outer\def\beginproclaim#1#2{\medbreak\noindent
                {\bf #1 \the\sectnum.#2.\enspace} \begingroup \sl}

\outer\def\endproclaim{\endgroup\par\medbreak}

\outer\def\beginremarks#1#2{\medbreak\noindent
                         {\bf #1 \the\sectnum.#2.\enspace}}
\outer\def\endremarks{\par \medbreak}

\outer\def\beginexample#1{\medbreak\noindent
                         {\bf Example \the\sectnum.#1.\enspace}}
\outer\def\endexample{\par \medbreak}

\outer\def\beginconjecture#1{\medbreak\noindent
                         {\bf Conjecture \the\sectnum.#1.\enspace}}
\outer\def\endconjecture{\par \medbreak}

\outer\def\beginproof{\medbreak\noindent {\it Proof.\enspace} }
\outer\def\endproof{\blackbox \par \medbreak}

\newcount\sectnum

%%% Use the following if equations cannot be automatically numbered
%%%
%%% \def\eqnum#1{\eqno {(\the\sectnum.#1)}}
%%% \def\eqalignnum#1{(\the\sectnum.#1)}

%%% Automatically numbered equations %%%
\newcount\num
\def\eqnum#1#2{\global\advance\num by 1
             \eqno {(\the\sectnum.\the\num {\rm #2})}
             \xdef#1{\the\sectnum.\the\num} }
\def\eqalignnum#1#2{\global\advance\num by 1
             {(\the\sectnum.\the\num {\rm #2})}
             \xdef#1{\the\sectnum.\the\num} }

\def\label#1#2{{(#1{\rm #2})}}

%%% Forward References %%%
\newwrite\ffile  %forward reference file
\newif\iffclosed \fclosedtrue % flag to indicate if fwd ref file
% opened yet
\def\fwrite#1{\iffclosed\immediate\openout\ffile=\jobname.fwd\fclo
sedfalse\fi
              \immediate\write\ffile{#1}}

\def\frwdref#1{\fwrite{\noexpand\xdef\noexpand#1\noexpand{#1}}}

\newread\tfile
\def\testinput#1{
                 \immediate\openin\tfile=#1

\immediate\ifeof\tfile\else\closein\tfile\input#1\fi
                 \closein\tfile
                  }
\testinput{\jobname.fwd}
%%%%%%%%%%%%%%%%%%%%%%%%%%%%%%%%%%%%%%%%%%%

\hsize=6.5in
\vsize=9in

%%%Page Break Parameters for Penalties
\widowpenalty=10000
\displaywidowpenalty=1000
\predisplaypenalty=5000
\postdisplaypenalty=100
\clubpenalty=10000

\magnification=1200

\pageno=1
{\largest
\centerline{Integrable highest weight modules over}
\centerline{affine superalgebras and number theory} }
\smallskip
\centerline{by Victor G. Kac$^1$\footnote{$^*$}{Supported in part
by NSF grant DMS-9103792.}
 and Minoru Wakimoto$^2$}
\centerline{Department of Mathematics, M.I.T., Cambridge, MA 02139}
\centerline{Department of Mathematics, Mie University, Tsu 514, Japan}
\vskip 2pt
\noindent
{\it To Bertram Kostant from whom we learned so much.}

\beginsection 0 Introduction

The problem of representing an integer as a sum of squares of
integers has had a long history.  One of the first after antiquity
was A.~Girard who in 1632 conjectured that an odd prime $ p\/$ can
be represented as a sum of two squares iff $ p \equiv 1 \mod 4\/$,
and P.~Fermat in 1641 gave an ``irrefutable proof'' of this
conjecture.  The subsequent work on this problem culminated in
papers by A.M.~Legendre (1798) and C.F.~Gauss (1801) who found
explicit formulas for the number of representations of an integer as
a sum of two squares.  C.G.~Bachet in 1621 conjectured that any
positive integer can be represented as a sum of four squares of
integers, and it took efforts of many mathematicians for about 150
years before J.-L.~Lagrange gave a proof of this conjecture in 1770.

A radically new approach to the problem, based on the theory of
elliptic functions, was found by C.G.J.~Jacobi in his famous {\it
Fundamenta nova\/} [J] published in 1829.  In order to state his
results, introduce the following power series in $ q\/$:
$$
\squaretwo (q) = \sum_{j \in \ZZ} q^{j^2}.
$$
Then, clearly, the coefficient of $ q^k\/$ in the power series
expansion of $ \squaretwo (q)^d\/$ is the number of representations
of $ k\/$ as a sum of $ d\/$ squares of integers (taking into
account the order of summands).  Jacobi discovered the following
remarkable identities, which gave a solution to the problem for $
d = $ 2, 4, 6 and 8:
$$
\leqalignno{
\squaretwo (q)^2 & = 1 + 4 \sum_{j,k=1}^\infty (-1)^k q^{j (2k -
   1)}, & (0.1) \cr
\squaretwo (q)^4 & = 1 + 8 \sum_{j,k=1 \atop 4 \nmid k}^\infty k
   q^{jk},
   & (0.2) \cr
\squaretwo (q)^6 & = 1 - 4 \sum^\infty_{j,k=1 \atop j>k, j-k {\rm
   odd}} (-1)^{{1\over2}(j-1)(k+1)} (j^2 - k^2) q^{{1 \over 2} jk},
   & (0.3) \cr
\squaretwo (q)^8 & = 1 + 16 \sum^\infty_{j,k=1} (-1)^{(j+1)k} k^3
   q^{jk}. & (0.4)
}% End of leqalignno
$$

Further development has been based, starting with the works of
Jacobi and Legendre, on the theory of modular forms.  Explicit
formulas for odd $ d \leq 7\/$ have been found by Legendre,
Eisenstein, Smith, Hardy and Ramanujan (the case of odd $ d\/$ is
much harder than that of even $ d\/$).  In the 1860s J.~Liouville
obtained a series of beautiful formulas for $ d = 10\/$ and 12.  In
1917 L.J.~Mordell showed how in principle the problem can be solved
for even $ d\/$ using the Fricke-Klein theory, but the difficulties
increase rapidly with $ d\/$.  One can find a detailed (but quite
disordered) account on sums of squares in Volume II of Dickson's
``History of the Theory of Numbers'' (Chapters VI--IX).

The related problem of representing a positive integer by sums of
triangular numbers ({\it i.e.,\/} numbers of the form $ {1 \over 2}
j ( j + 1)\/$, $ j = 0, 1, \ldots \/$) has also been quite popular,
and a number of deep results was obtained by Gauss, Legendre and
others (see Chapter 1 of Vol.~2 of Dickson's ``History'').  Here the
problem, clearly, is to find a power series expansion of powers of
the series
$$
\Delta (q) = \sum^\infty_{j = 0} q^{{1 \over 2} j (j+1)} .
$$

The main objective of the present paper is to demonstrate that
representation theory of affine superalgebras is a natural framework
for the study of the problem of sums of squares and sums of
triangular numbers.  However, in order to state our main result we
need only the following notions.  Let $ V\/$ be a finite-dimensional
vector space over $ \RR\/$ with a non-degenerate symmetric bilinear
form $ (.|.)\/$.  Let $ \Delta\/$ be a finite set of vectors in $
V\/$ not containing 0, called the set of {\it roots\/}, and let $
\Delta = \Delta_0 \cup \Delta_1\/$ be represented as a disjoint
union of two subsets, called the sets of {\it even\/} and {\it odd
roots\/}.  A choice of a linear function $ f\/$ on $ V\/$ which does
not vanish on $ \Delta\/$ defines a subset of {\it positive roots\/}
$ \Delta_+ = \{ \alpha \in \Delta | f(\alpha) > 0 \}\/$; we let $
\Delta_{\epsilon+} = \Delta_\epsilon \cap \Delta_+\/$, $ \epsilon =
0, 1\/$.  Given $ \alpha \in V\/$ we define the exponential function
$ e^\alpha\/$ by $ e^\alpha (v) = e^{(\alpha | v)}\/$.  We call the
{\it Weyl denominator\/} associated to the above {\it root data\/}
the function
$$
\R =
\left.
   \prod_{\alpha \in \Delta_{0+}} ( 1 - e^{-\alpha})
\right/
   \prod_{\alpha \in \Delta_{1+}} (1 + e^{-\alpha}) .
$$
Let $ q\/$ be an indeterminate; we call the {\it affine
denominator\/} the infinite product
$$
\hat\R = \R
\prod^\infty_{n=1}
\left(
   (1 - q^n)^\ell
   \left.
      \prod_{\alpha \in \Delta_0} (1 - q^n e^\alpha)
   \right/
\prod_{\alpha \in \Delta_1} (1 + q^n e^\alpha)
\right) ,
$$
where $ \ell\/$ is the dimension of the $ \RR\/$-span of $
\Delta\/$.  Our main objective is to find expansions for R and $
\hat\R\/$.  Of course, in the case when $ \Delta = \Delta_0\/$ is
the set of roots of a simple finite-dimensional Lie algebra, an
expansion for R (resp.\ $ \hat\R\/$) is given by the well-known Weyl
denominator identity (resp.\ Macdonald identity [Mac]), see {\it
e.g.,\/} [K9].  Our main Theorems 2.1 and 4.1 extend these
results to the case when $ \Delta\/$ is a set of roots of a finite-
dimensional simple Lie superalgebra $ \gg\/$ with a non-degenerate
Killing form.

The connection to sums of squares is obvious.  Indeed
$$
\left.
   {\widehat\R \over \R} (0) = \prod^\infty_{n=1} (1 - q^n)^{\ell
   + |\Delta_0|}
\right/
(1 + q^n)^{|\Delta_1|},
$$
and if it happens that
$$
\ell + |\Delta_0| = |\Delta_1| =: d ,
\leqno(0.5)
$$
then we get an expansion of $ \squaretwo (q)^d\/$, in view of Gauss'
identity
$$
\squaretwo (q) = \prod^\infty_{n=1} {1 - q^n \over 1 + q^n} .
$$
Evaluating at points other than 0, one gets expansions for
triangular numbers (and, more generally, $ m\/$-gonal numbers).

To be more specific, let us give an example of a root system of
type $ A (m - 1, n - 1)\/$, where $ m > n \geq 1\/$.  Let $ V\/$
be the $ (m+n)\/$-dimensional vector space over $ \RR\/$ with
basis $ \epsilon_1, \ldots, \epsilon_m\/$, $ \delta_1, \ldots,
\delta_n\/$ and the symmetric bilinear form given by $
(\epsilon_i | \epsilon_i) = - ( \delta_j | \delta_j) = 1\/$ for
all {\it i\/} and {\it j\/} and zero for any other pair.  Let $
\Delta_0 = \{ \epsilon_i - \epsilon_j | 1 \leq i\/$, $ j \leq
m\/$, $i \neq j \} \cup \{ \delta_i - \delta_j | 1 \leq i\/$, $ j
\leq n\/$, $i \neq j\}\/$, $ \Delta_1 = \{ \epsilon_i - \delta_j,
\delta_j - \epsilon_i | 1 \leq i \leq m, 1 \leq j \leq n \}\/$, $
\Delta = \Delta_0 \cup \Delta_1\/$.  Let $ \Delta_+ = \{
\epsilon_i - \epsilon_j, \delta_i - \delta_j | i < j \} \cup \{
\epsilon_i - \delta_j | i \leq j \} \cup \{ \delta_i - \epsilon_j
| i < j \}\/$.  Condition (0.5) holds iff $ m = n + 1\/$.  In
this case, {\it i.e.,\/} $ \Delta\/$ of type $ A (n, n-1)\/$, the
denominator identity is
$$
\matrix{
  \displaystyle
  \R_n := \prod_{i<j}
  \left(
     1 - {x_i \over x_j}
  \right)
  \prod_{i < j}
  \left(
     1 - {y_i \over y_j}
  \right)
  \prod_{i \leq j}
  \left(
     1 + {x_i \over y_j}
  \right)^{-1}
  \prod_{i < j}
  \left(
     1 + {y_i \over x_j}
  \right)^{-1} \cr
  \displaystyle
  = \sum_{w \in S_{n+1}} \epsilon (w) \prod^n_{i=1}
  \left(
     1 + {y_i \over x_{w(i)}}
  \right)^{-1} . \qquad\qquad & \cr
}% End of eqalign
\leqno(0.6)
$$
Here $ x_i = e^{-\epsilon_i}\/$, $ y_i = e^{-\delta_i}\/$, $ i\/$
and $ j\/$ take all possible values and $ \epsilon(w)\/$ is the sign
of the permutation $ w \in S_{n+1}\/$.  Furthermore, the affine
denominator identity reads:
$$
\matrix{\displaystyle
  \R_n \prod^\infty_{k=1}
  \left(
     (1 - q^k)^{-1}
     \left.
        \prod_{1 \leq i, j \leq n + 1 \atop 1 \leq r, s \leq n}
        \left(
           1 - q^k {x_i \over x_j}
        \right)
        \left(
           1 - q^k {y_r \over y_s}
        \right)
     \right/
     \left(
        1 + q^k {x_i \over y_r}
     \right)
     \left(
        1 + q^k {y_s \over x_j}
     \right)
  \right)
  \cr
  \displaystyle
  = \sum_{w \in S_{n+1}} \epsilon(w) \sum_{\vec m \in M_n}
      q^{{1 \over 2} \sum^{n+1}_{i=1} m^2_i} \vec x^{\vec m}
      \prod_{i=1}^n
      \left(
         1 + {y_i \over x_{\sigma(i)}} q^{-m_i}
      \right)^{-1} . \cr
}% end of eqalign
\leqno(0.7)
$$
Here $ M_n\/$ is the sublattice of the lattice $ \sum_i \ZZ
\epsilon_i\/$ of vectors with zero sum of coordinates (= the root
lattice of type $ A_n\/$) and $ \vec x^{\vec m} := x^{m_1}_1 \ldots
x^{m_{n+1}}_{n+1}\/$.  Dividing both sides of (0.7) by $ \R\/$ and
taking limit as $ x_i\/$, $ y_i \rightarrow 1\/$, we obtain an
expansion of $ \squaretwo (q)^{2n (n + 1)}\/$.  One of the ways to
actually compute this limit is explained in the proof of Theorem
4.2, but probably there is a different way giving a better formula.
Of course, for $ n = 1\/$ we thus recover Jacobi's formula (0.2).
Another specialization of (0.7) gives the following expansion (see
Example~4.3b):
$$
\squaretwo (-q)^{2n} = 2^n \sum_{\vec m \in M_n}
\sum_{s \in \SS} (-1)^{|s|}
q^{{n + 1 \over 2} \sum^{n+1}_{i=1} (m^2_i + m_i - m_i \tilde s_i)}
\prod^n_{j=1}
\left(
   1 + q^{(n+1) (m_j - \tilde s_j)}
\right)^{-1} ,
\leqno(0.8)
$$
where $ \SS\/$ is the set of all subsets of the set $ \{ \epsilon_i
- \epsilon_j | i < j \}\/$ (= set of positive roots of type $
A_n\/$), and $ \tilde s\/$ (resp.\ $ |s|\/$) denote the sum (resp.\
number) elements from $ s \in \SS\/$.  Formula (0.8) for $ n = 1\/$
is an equivalent form of (0.1).

Even the simplest case of (0.6), when $ n = 1\/$, is a truly
remarkable identity.  Letting $ u = -x_1 / y_1\/$, $ v = -x_2 /
y_1\/$, it takes the following beautiful symmetric form:
$$
\prod_{n=1}^\infty
{(1 - q^n)^2 (1 - uv q^{n-1}) (1 - u^{-1} v^{-1} q^n)
\over
(1 - u q^{n-1}) (1 - u^{-1} q^n) (1 - v q^{n-1}) (1 - v^{-1} q^n)}
=
\left(
   \sum^\infty_{m,n=0} - \sum^{-\infty}_{m,n=-1}
\right)
u^m v^n q^{mn}.
\leqno(0.9)
$$
After establishing this identity, we realized that it has been
rediscovered many times in the past.  It is implicitly contained in
[KL] as it follows immediately by comparing characters in the super
boson-fermion correspondence (which gives a combinatorial proof of
(0.9)), and in [M] where it is used to rewrite the characters of $
N = 2\/$ superconformal algebras in a multiplicative form.  As has
been pointed out in [F] and [H], (0.9) is a special case of the
Ramanujan summation formula for the bilateral basic hypergeometric
function $ {}_1 \Psi_1\/$.

Condition (0.5) holds iff $ \gg\/$ is of type $ A (n, n-1)\/$, $
B (n, n)\/$ or $ D (n + 1, n + 1)\/$ ($ n = 1, 2, \ldots\/$),
which gives an expansion for $ \squaretwo (q)^N\/$ for $ N = 2n
(n + 1)\/$, $ 2n (2n + 1)\/$ and $ 4 (n+1)^2\/$ respectively
(given by Theorem 4.2; see also Examples 5.3 and 5.4).  The first
members of the first and the second series are Jacobi's formulas
(0.2) and (0.3).  The second member of the first series gives an
expansion for $ \squaretwo (q)^{12}\/$, however we do not know
how to derive from this expansion the famous result of
J.~Liouville (1860) that the number of representations of $
2k\/$, where $ k\/$ is odd, as a sum of 12 squares equals $ 264
\sum_{j|k} j^5\/$.  Finally note that Jacobi's formula (0.4) is
obtained by evaluating at 0 the (conjectured) denominator
identity for the case $ \gg = gl (2,2)\/$, which is a Lie
superalgebra with a zero Killing form (see Conjecture~7.1).
Another very interesting case which is not covered by Theorem 4.1
is $ \gg = Q (m)\/$ (see Conjecture~7.2).  A specialization of
the corresponding affine denominators gives very beautiful power
series expansions of $ \Delta (q)^{4 s^2}\/$ and $ \Delta (q)^{4
s (s + 1)}\/$ (see Section 7).  For $ s = 1\/$ we thus recover
Legendre's formulas for sums of 4 and 8 triangular numbers
(whereas formulas for 2, 4 and 6 triangular numbers are obtained
by certain specializations of the affine denominator identities
for $ A (1,0)\/$, $ A (1,0)\/$ and $ B(1,1)\/$ respectively, see
Section 5). For $ s \geq 2\/$ we obtain formulas which are
probably new.  For example, we find that the number of
representations of $ n\/$ as a sum of 16 triangular numbers
equals
$$
{1 \over 3 \cdot 4^3}
\sum_{{a,b,r,s \in \NN {\rm odd} \atop a > b}
      \atop ar + bs = 2n + 4}
ab (a^2 - b^2)^2 .
$$

In Section 1 we recall the necessary material on a simple
finite-dimensional Lie superalgebra $ \gg\/$ [K2].  The new material
here concerns the dual Coxeter number $ h^\vee\/$ (which is the
eigenvalue of the Casimir operator associated to the suitably
normalized Killing form).  As in the Lie algebra case, $ h^\vee\/$
plays an important role in the theory of affine superalgebras.  In
Section 2 we introduce the important notion of the defect of $
\gg\/$ (which is 0 in the Lie algebra case) and of a maximal
isotropic set of roots.  This allows us to state the denominator
identity (Theorem 2.1) for the root system of $ \gg\/$.  We give a
proof which works in the case of $ \mydef \gg \leq 1\/$ (which takes
care of all exceptional Lie superalgebras).  A proof in the general
case involves more calculations and will be published elsewhere.  At
any rate, the proof is based on the analysis of irreducible
subquotients of the Verma module $ M (0)\/$ over $ \gg\/$ with
highest
weight 0.  Since the denominator identity may be viewed as the
character of the 1-dimensional $ \gg\/$-module, it is only natural
to go on and try to compute in Section 3 the character of an
arbitrary finite-dimensional irreducible $ \gg\/$-module $ V\/$.
There has been a number of papers concerned with this problem [K3,
K4, BL, TM1, TM2, VJ1, VJ2, JHKT1, JHKT2, P, PS1, PS2, S, Se,
$\ldots$].
Our basic new ingredient is the notion of a $ \lambda\/$-maximal
isotropic subset, which allows us to define atypicality of $ V\/$
which is independent of the choice of the set of positive roots, and
of a tame $ \gg\/$-module $ V\/$.  Our first basic result is
Theorem~3.1 which states that $ V\/$ is tame (and hence gives a
formula for $ \ch V\/$) in the case when for some choice of the set
of positive roots there exists a $ \Lambda + \rho\/$-maximal
isotropic subset consisting of simple roots.  (Note that Theorem~2.1
is a special case of Theorem~3.1 since $ \mydef \gg = \atp 0\/$.)
This theorem covers most of the known formulas for $ \ch V\/$
(proved in [K3, K4, BL, VJ1, VJ2, JHKT2, $\ldots$]), but
unfortunately there
exist $ V\/$ that are not tame.  We describe all cases of non-tame
$ V\/$ of atypicality $ \leq 1\/$; this gives, in particular,
character formulas for all $ V\/$ over exceptional Lie superalgebras
(Example~3.3).  We also give a proof of a dimension formula
(Theorem~3.2), which is based on a nice regularization procedure,
and of a superdimension formula (Theorem~3.3), which is based on an
interesting property of the root system of $ \gg\/$ (Lemma~2.2d) and
on Theorem~2.1.

In the second part of the paper (Sections 4--8) we turn to the
associated to $ \gg\/$ affine superalgebra $ \hat\gg\/$.  The
main result is Theorem~4.1 which gives the affine denominator
identity in the case when the Killing form is nondegenerate
(equivalently, $ h^\vee \not= 0\/$).  The proof is again based on
the analysis of irreducible subquotients of the Verma module over
$ \hat\gg\/$ with zero highest weight.  (Of course, this is the
basic idea of the proof of Macdonald identities given in [K1].)
Examples~4.4 and 4.5 give a more convenient form for
specializations of the affine denominator identity.  Theorems~4.2
and 4.3 and Example~4.3b give explicit formulas for two
specializations of this identity.  These formulas are applied in
Section~5 to derive expressions for the number of representations
of an integer as sums of squares and sums of triangular numbers.
In Section~6 we use the simplest affine denominator identity to
study transformation properties of certain ``degenerate'' theta
functions and modular forms and apply this to prove
transformation properties of the $ N = 2\/$ superconformal
algebras, {\it cf.\/} [RY].  (These results may be extended to
the general case by the same method.)  In Section~7 we discuss
the affine denominator identity for $ \gg = gl (2,2)\/$, the
simplest case when $ h^\vee = 0\/$.  Another extremely
interesting case is $ \gg = Q(m)\/$.  We conjecture the
corresponding affine denominator identity and derive formulas for
the number of representations as sums of triangular numbers in
terms of dimensions of irreducible representations of $
sl_{m+1}\/$.  Finally, in Section~8 we introduce the notion of an
integrable $ \hat\gg\/$-module, based on our experience with the
1-dimensional module.  Theorem~8.1 gives a classification of
irreducible integrable highest weight $ \hat\gg\/$-modules $ \hat
L(\Lambda)\/$.

(As in the Lie algebra case, the level of such a module is a
non-zero integer, but there may be infinitely many such modules of
positive level.)  It is an extremely interesting problem to
calculate the characters of integrable $ \hat L (\Lambda)\/$.  A
partial answer is given for $ A (1,0)^\wedge\/$ (Example~8.1).  Of
course, in the case when $ \mydef \gg = 0\/$ the answer is well
known for both $ \gg\/$ and $ \hat\gg\/$ [K5].

\bigskip
\noindent{\bf Acknowledgment}\medskip

We learned from V.~Serganova the very useful method of odd
reflections.  Discussions with her and I.~Penkov were very
illuminating.  We relied extensively on the (unpublished) tables of
J.~Thierry-Mieg [TM1] and computer calculations of J.~Van der Jeugt of
dimensions and superdimensions of irreducible $ \gg\/$-modules.
Discussions with them and R.C.~King were very useful.  We are very
grateful to all these people.

The results of this work were reported at the conference on
universal enveloping algebras in Paris in June 1992 and at the
Kostant conference in May 1993.

\beginsection 1 Dual Coxeter number

Let $\gg = \gg_0 + \gg_1\/$ be a finite-dimensional simple Lie
superalgebra
over ${\Bbb C}\/$ with a non-degenerate even invariant bilinear form
$B\/$.
Recall that ``even'' (resp.~``invariant'') means that
$B (\gg_0, \gg_1) = 0\/$ (resp.~$B
\bigl( [a,b],c \bigr) = B \bigl( a,[b,c] \bigr), a, b, c \in \gg\/$)
and that all such forms are proportional.  Denote by
$h^\vee_B\/$ the eigenvalue of the Casimir operator $\Omega_B\/$
associated to $B\/$ [K2, p.~85] in the adjoint representation.

\beginproclaim {Proposition} {1}
$h^\vee_B \neq 0\/$ if and only if the Killing form is
non-degenerate.
\endproclaim

Choose a Cartan subalgebra $\hh\/$ of $\gg_0\/$, let $\Delta \in
\hh^\ast\/$ be the set of roots, $\Delta_0\/$ and $\Delta_1\/$ the
subsets of even and odd roots and let $W \subset G L (\hh^\ast)\/$
be the Weyl group (= the group generated by reflections $r_\alpha\/$
with respect to even roots $\alpha\/$).  Let $\gg = \hh \oplus
\Bigl( \relbuilder {\oplus}{\alpha \in \Delta} \gg_\alpha \Bigr)\/$
be the root space decomposition.

Recall that an odd root $\alpha\/$ is isotropic
$\bigl(\/$i.e.\ $B(\alpha, \alpha) = 0 \bigr)\/$ iff $2 \alpha\/$
is not a root.  We let
$$
\bar{\Delta}_0 = \bigl\{ \alpha \in \Delta_0 \bigm|
     {\textstyle {1 \over 2}} \alpha \notin  \Delta \bigr\}, \quad
\bar{\Delta}_1 = \bigl\{ \alpha \in \Delta_1 \bigm| (\alpha |
\alpha) = 0\bigr\}.
$$
Then $\Delta_1 \setminus \bar{\Delta}_1 = {1 \over 2}
(\Delta_0 \backslash \bar{\Delta}_0)\/$ and
$\bar{\Delta}_0\/$ is a root system.

Choose a set of positive roots $\Delta_+\/$ in $\Delta\/$, let
$\Delta_{\epsilon +} = \Delta_\epsilon \cap \Delta_+\/$ ($\epsilon
= 0\/$ or 1) and let  $\nn_+ = \relbuilder {\oplus}{\alpha \in
\Delta_+} \gg_\alpha\/$.  Let $\Pi = \{ \alpha_1, \ldots, \alpha_l
\} \subset \Delta_+\/$ be the set of simple roots and let $\theta
\in \Delta_+\/$ be the highest root.  Let $\rho_\epsilon\/$
be the half of the sum of the roots from $\Delta_{\epsilon +}\/$ and
and let $\rho = \rho_0 - \rho_1\/$.  Then, $B(\rho, \alpha_i) = {1
\over 2} B (\alpha_i, \alpha_i)\/$, $i = 1, \ldots, l\/$.
Furthermore one has
$$
h^\vee_B = B (\rho, \rho) + {\textstyle {1 \over 2}} B
(\theta, \theta).
\eqno(1.1)
$$

The following generalization of Freudenthal-de~Vries ``strange
formula'' holds:
$$
B (\rho, \rho) = {{h^\vee_B} \over {12}} (\dim \gg_0 - \dim \gg_1).
\eqno(1.2)
$$

Assuming that $h^\vee_B \neq 0\/$, let
$$
\Delta^\sharp_0 = \bigl\{ \alpha \in \Delta_0 \bigm|
              h^\vee_B B(\alpha, \alpha) > 0 \bigr\}
$$
(this is independent of the choice of $B\/$), and let $W^\sharp\/$
denote the  subgroup of $W$ generated by reflection $r_\alpha\/$
with respect to all $\alpha \in \Delta^\sharp_0\/$.  Denote by
$(.|.)$ the even invariant bilinear form on $\gg\/$ normalized by
the condition
$$
(\alpha | \alpha) = 2 \hbox{ for the longest root } \alpha \in
\Delta^\sharp_0.
\eqno(1.3)
$$
The corresponding to this form number $h^\vee = h^\vee_{(.|.)}\/$ is
called the {\it dual Coxeter number\/} of the Lie superalgebra
$\gg\/$.  We let $h^\vee = 0\/$ if $h^\vee_B = 0\/$.  We give below
the number $h^\vee\/$ in all cases (excluding Lie algebra cases):
$$
\eqalign {
   A (m,n)  \, : \; & \; h^\vee = | m - n | ; \quad
                    C (n) : h^\vee = n - 1; \cr
   B (m,n)  \, : \; & \; h^\vee = 2 (m - n) - 1 \hbox{ if } m > n,
      \quad
                     = n - m + {\textstyle {1 \over 2}} \hbox{
                     if } m \leq n; \cr
   D (m,n)  \, : \; & \; h^\vee = 2 (m - n - 1) \hbox{ if } m \geq n
      + 1, \quad
                     = n - m + 1 \hbox{ if } m < n + 1;
   \cr
      D (2,1; \alpha) \, : \; & \; h^\vee = 0; \quad
                                F (4) : h^\vee = 3; \quad
                                G (3) : h^\vee = 2. \cr
}% End of eqalign
$$

\beginremarks{Remarks}{1}
(a) Let $\kappa (\ .\ ,\ .\ )\/$ denote the Killing form, which we
assume to be non-degenerate.  Then $\Delta^\sharp_0 = \bigl\{ \alpha
\in \Delta_0 \bigm| \kappa (\alpha, \alpha) > 0  \bigr\} \/$ is the
root system of one of the simple components of $\gg_0\/$ and $h^\vee
= \kappa (\alpha, \alpha)^{-1}\/$ for the longest $\alpha \in
\Delta^\sharp_0\/$.

(b) $h^\vee = 0\/$ if and only if $\gg\/$ is of type $A (n, n)\/$,
$D (n + 1, n)\/$ or $D (2, 1; \alpha)\/$ [~~~].  In these cases we
define $\Delta^\sharp_0\/$ as follows.  The root system $\Delta_0\/$
is a union of two orthogonal to each other root subsystem; namely
one has respectively:
$$
\Delta_0 = A_n \cup A_n, \quad
D_{n + 1} \cup C_n, \quad
D_2 \cup C_1.
$$
Then we let $\Delta^\sharp_0\/$ to be the first subset and
$W^\sharp\/$ to be the subgroup of $W$ generated by the $r_\alpha\/$
with $\alpha \in \Delta^\sharp_0\/$.  We normalize $(.|.)$ by
(1.3) as before.

(c) There are many $ W\/$-inequivalent choices of $ \Delta_+\/$.
However, there is a unique (up to an automorphism of the root
lattice) choice of $ \Delta_+\/$ such that $ \Pi\/$ contains a
unique odd root; we call this $ \Delta_+\/$ {\it standard\/}.
\endremarks

In order to prove the above statements, recall that [K2, p.~85]
the eigenvalue of the Casimir operator $\Omega_B\/$ in a
representation with highest weight $\Lambda\/$ is equal to $B
(\Lambda, \Lambda + 2 \rho)\/$ and that when the Killing form
$\kappa\/$ is non-degenerate one has:
$$
\kappa (\theta, \theta + 2 \rho) = 1
\eqno(1.4)
$$
provided that $\dim \gg_0 \neq \dim \gg_1\/$.  Since
$$
\dim \gg_0 = \dim \gg_1 \quad \hbox{iff} \quad
\gg \simeq A (n + 1, n), \;
B (n, n) \hbox{ or } D (n, n)
\eqno(1.5)
$$
and since (1.4) is a polynomial identity which holds
for all other values of $(m, n)\/$ (except when $\kappa \equiv 0\/$)
we obtain that (1.4) holds in all cases when $\kappa\/$ is
non-degenerate.  This proves Proposition 1.1 and formula (1.1).

{}From formula (2.1) in \S 2, it follows that
$$
B (\rho, \rho) \hbox{ is independent of the choice of } \Delta_+.
\eqno(1.6)
$$
Hence it suffices to check (1.2) for one choice of $\Delta_+\/$.  A
more conceptual proof of (1.2) is based on the theory of
``degenerate'' theta functions ({\it cf.\/} Remark 6.1 in Section
6).

\beginexample{1}
$ \gg = A (m, n)\/$ with $ m > n\/$.  Choose $ \Delta_+\/$ such that
the number of odd simple roots is maximal.  Then $ \Pi = \Pi' \cup
\Pi''\/$, where $ (\alpha | \alpha ) = 0\/$ if $ \alpha \in \Pi'\/$
and $ (\alpha | \alpha) = 2\/$ if $ \alpha \in \Pi''\/$, and one can
order the sets $ \Pi'\/$ and $ \Pi''\/$:
$$
\Pi' =
\left\{
   \alpha_1, \alpha_2, \ldots, \alpha_{2 (n+1)}
\right\}, \quad
\Pi'' =
\left\{
   \alpha_{2n+3}, \ldots, \alpha_{m+n+1}
\right\}
$$
in such a way that all non-zero scalar products of distinct roots
are:
$$
\eqalign{
(\alpha_{2j-1} | \alpha_{2j}) & = 1 \hbox{ if } 1 \leq j \leq n +
      1,\quad
      (\alpha_{2j} | \alpha_{2j+1}) = -1 \hbox{ if } 1 \leq j \leq
      n, \cr
      (\alpha_j | \alpha_{j+1}) & = -1 \hbox{ if } 2n + 2 \leq j \leq
      m + n. \cr
}% End of eqalign
$$
In particular, $ \Pi'\/$ (resp.\ $ \Pi''\/$) is a set of simple
roots of $ A (n + 1, n)\/$ (resp.\ $ A_{m-n-1}\/$).  The root system
$ \Delta^\sharp_0\/$ is of type $ A_{m}\/$ and has the following
set of simple roots:  $ \{ \alpha_1 + \alpha_2, \ldots,
\alpha_{2n+1} + \alpha_{2n+2}, \alpha_{2n+3}, \ldots,
\alpha_{m+n+1}\}\/$.  In the (most important for us) case when $ m =
n + 1\/$, we have $ \Pi = \Pi'\/$ and $ \Delta^\sharp_0\/$ has
simple roots $ \{ \alpha_1 + \alpha_2, \ldots, \alpha_{2n+1} +
\alpha_{2n+2} \}\/$.  Note that (1.2) for $ A(m,n)\/$ follows from
the ``strange formula'' for $ A_{m-n-1}\/$.
\endexample

\beginsection 2 The Weyl denominator of $\gg\/$

Given a subset of positive roots $\Delta_+\/$ of $\Delta\/$ and a
simple root
$\alpha\/$, we may construct a new subset of positive roots
({\it cf.}~[PS2]):
$$
\eqalign{
     \Delta'_+ = \; & r_\alpha (\Delta_+) \hbox{~~if~~}
                    (\alpha | \alpha) \neq 0,        \cr
     \Delta'_+ = \; & \bigl( \Delta_+ \cup \{ - \alpha \} \bigr)
                    \setminus \{ \alpha \} \hbox{~~if~~}
                    (\alpha | \alpha) = 0.           \cr
}
$$
The set $\Delta'_+\/$ is called a simple reflection of $\Delta_+\/$.
Then one has:
$$
\rho' =  \left\{
                   \matrix{ \rho - \alpha & \hbox{if} &
                                (\alpha | \alpha) \neq 0   \cr
                            \rho + \alpha & \hbox{if} &
                                (\alpha | \alpha) = 0.     \cr
}% End of matrix
\right.
\leqno(2.1)
$$
Note that $\Delta'_+\/$ and $\Delta_+\/$ are $W$-inequivalent
if $(\alpha | \alpha) = 0\/$.  By a usual argument, any two subsets
of positive roots of $\Delta\/$ may be obtained from each other by
a sequence of simple reflections.

By a {\it regular exponential function\/} on $\hh\/$ we mean a
finite linear combination of exponentials $e^\lambda\/$, $\lambda
\in \hh^*\/$.  {\it A rational exponential\/} function is a ratio $P
/ Q\/$ where $P\/$ and $Q\/$ are regular exponential functions and
$Q \neq 0\/$.  The Weyl group $W$ acts on the field of rational
exponential functions via $e^\lambda \mapsto e^{w (\lambda)}\/$, $w
\in W$.

The rational exponential function
$$
R = \left. \prod_{\alpha \in \Delta_{0 +}} (1 - e^{- \alpha})
 \right/ \prod_{\alpha \in \Delta_{1 +}} (1 + e^{- \alpha})
\leqno(2.2)
$$
is called the {\it Weyl denominator\/} of $\gg\/$.  Introduce also
the {\it Weyl superdenominator\/}:
$$
\eqalign{
\check{R}
   & = \left. \prod_{\alpha \in \Delta_{0 +}}
            (1 - e^{- \alpha}) \right/
            \prod_{\alpha \in \Delta_{1 +}} (1 - e^{- \alpha})  \cr
   & = \prod_{\alpha \in \bar{\Delta}_{0 +}}
                         (1 - e^{- \alpha})
                   \left.\prod_{\alpha \in \Delta_{1 +} \setminus
                                   \bar{\Delta}_{1 +}}
                         (1 + e^{- \alpha})  \right/
                   \prod_{\alpha \in \bar{\Delta}_{1 +}}
                         (1 - e^{- \alpha}) .    \cr
}% End of eqalign
\leqno(2.3)
$$

The proof of the following lemma is straightforward.

\beginproclaim {Lemma} {1}
Let $\Delta_+\/$ and $\Delta'_+\/$ be two subsets of positive roots
in $\Delta\/$, and let $R\/$ and $R'\/$ (resp.~$\check{R}$ and
$\check{R}'\/$) be the corresponding Weyl (resp.\ super)
denominators.  Then
$$
e^\rho R = (- 1)^{|\Delta_{0 +} \cap \Delta'_{0 +}|} e^{\rho'} R',
\quad
e^\rho \check{R} = (-1)^{|\bar{\Delta}_{0 +} \cap \bar{\Delta}'_{0
+}|} e^{\rho'} \check{R}'.
$$
\endproclaim

The basic difference between the Lie algebra and superalgebra cases
is that the restriction of the bilinear form $(.|.)$ to the subspace
$V := \Sum_{\alpha \in \Delta} \Bbb R \alpha\/$ may be indefinite;
the dimension of a maximal isotropic subspace of $V\/$ is called the
{\it defect\/} of $\gg\/$ and is denoted by $ \mydef \gg\/$.  Here
is a list of defects
$$
\eqalign{
   A (m-1, n-1), \;\; B (m, n), \;\; D (m, n) & \; :\; \mydef \gg
      = \min (m, n)  \cr
   C (n), D (2, 1; \alpha), F (4), G (3) & \; : \; \mydef \gg = 1.
      \cr
}% End of eqalign
$$
(Note that $\mydef \gg = 0\/$ iff $\bar{\gg}\/$ is a Lie
algebra or a Lie superalgebra $B (0, n)\/$.)

A subset $S\/$ of $\Delta\/$ is called {\it maximal isotropic\/} if
it consists of $ \mydef \gg\/$ roots that span a maximal isotropic
subspace of~$V\/$.  The following can be checked by a
straightforward calculation.

\beginproclaim {Lemma} {2}
(a) A maximal isotropic set of roots always exists.

(b) Given a maximal isotropic subset of roots $S\/$ there exists
$\Delta_+\/$ such that $S \subset \Pi\/$.

(c) Given two maximal isotropic subsets of roots $S = \{ \beta_1,
\ldots, \beta_s \}\/$ and $S' = $ \break $\{ \beta'_1, \ldots,
\beta'_s \}\/$, there exists $w \in W^\sharp\/$ such that $w
(\beta'_i) = \pm \beta_{\sigma(i)}\/$ for a permutation
$\sigma\/$.

(d) Let $S \subset \bar{\Delta}_{1+}\/$ be a maximal isotropic
subset and let $\bar{\Delta}^S_{0+} =  \{ \alpha \in \bar{\Delta}_{0
+} | \alpha \bot S \}\/$. Then
$$
                 | \bar{\Delta}_{1+} \setminus S | =
                 | \bar{\Delta}_{0+} \setminus
                         \bar{\Delta}^S_{0+} |.
$$
\endproclaim

\beginproclaim {Theorem} {1}
Let $S\/$ be a maximal isotropic subset of roots and choose
$\Delta_+\/$ such that $S \subset \Pi\/$.  Then
$$
e^\rho R = \Sum_{w \in W^\sharp} \epsilon (w) w
{e^\rho \over \prod_{\beta \in S} (1 + e^{- \beta})}
\leqno(2.4)
$$
\endproclaim

This theorem is a special case of Theorem 3.1 in Section 3.

Here we shall give a proof in the case $ \mydef \gg = 1\/$.  The
proof is based on the following

\beginproclaim{Lemma}{3}
Suppose that an irreducible $ \gg$-module with highest weight $
\lambda \in \hh^\ast$ is a subquotient of a Verma $ \gg$-module $ M
(\Lambda)$.  Then there exists a sequence of weights $ \lambda_0 =
\Lambda$, $ \lambda_1, \ldots, \lambda_k = \lambda$ of $ M
(\Lambda)$ such that $ \lambda_i$ ($ 0 < i \leq k$) is obtained
from $ \lambda_{i-1}$ in one of the following two ways:
\itemitem{(i)}
   $ \lambda_i = \lambda_{i-1} - n \gamma_i$, where $ \gamma_i
   \in \Delta_{0+}$ and $ (\lambda_{i-1} + \rho | \gamma^\vee_i)
   = n \in \NN$,

\itemitem{(ii)}
   $ \lambda_i = \lambda_{i-1} - \gamma_i$, where $ \gamma_i \in
   \bar\Delta_{1+}$ and $ (\lambda_{i-1} + \rho | \gamma_i ) =
   0$.
\endproclaim

\beginproof
See [K4, Theorem 3b].  (Note that the exponents in the
determinant formula given by [K4, Theorem 3a] should be corrected
as in [K8].)
\endproof

\medbreak\noindent
{\bf Proof of Theorem 2.1\enspace}
(case $ \mydef \gg = 1$.)  As usual, we have:
$$
e^\rho \R = \sum_{\lambda \in F} c_\lambda e^{\lambda + \rho}, \quad
c_\lambda \in \ZZ, \quad
c_\lambda \neq 0,
\leqno(2.5)
$$
where $ F\/$ is the set of weights described by (i) and (ii) in
Lemma~2.3 for $ \Lambda = 0\/$.  Choose $ \Delta_+\/$ such that $
\Pi\/$ contains a unique odd (isotropic) root, say, $ \beta\/$.  We
clearly have:
$$
\sum_{j \geq 0} c_{-j \beta} e^{\rho - j \beta} =
{e^\rho \over 1 + e^{-\beta}}.
\leqno(2.6)
$$
Consider the minimal $ i\/$ for which the possibility (ii) of
Lemma~2.3 might occur.  Then $ \lambda_{i-1} = w (\rho) -
\rho\/$, where $ w = r_{\gamma_{i - 1}} \ldots r_{\gamma_1} \in
W\/$ and $ \lambda_i = \lambda_{i-1} - \gamma_i\/$, $ \left(\rho |
w^{-1} (\gamma_i) \right) = 0\/$, $ \gamma_i \in \bar\Delta_1\/$.
But by a case-wise inspection, $ (\rho | \gamma) = 0\/$, $ \gamma
\in \bar\Delta_1 \Rightarrow \gamma = \pm \beta\/$.  Hence $
\gamma_i = w (\beta)\/$.  Consider now the minimal $ j \leq i - 1\/$
for which $ \gamma_j \not\in \Delta^\sharp_+\/$, {\it i.e.}, $
\gamma_j = \theta\/$.  But $ (\rho | \theta^\vee ) < 0\/$, hence we
conclude that $ w \in W^\sharp\/$.  Thus, $ \lambda_i = w (\rho -
\beta) - \rho\/$, $ w \in W^\sharp\/$.

Furthermore, $ (\beta | \theta^\vee) \geq 0\/$, hence $ (\rho -
\beta | \theta^\vee) < 0\/$, hence if $ s\/$ is minimal $ > i\/$ for
which possibility (ii) of Lemma~2.3 might occur, we conclude that $
\lambda_{s-1} = w_1 (\rho - \beta) - \rho\/$, $ w_1 \in W^\sharp\/$.

By a case-wise inspection, we see that $(\rho - \beta | \beta_1 ) =
0$, $\beta_1 \in \bar\Delta_{1+} \Rightarrow \beta_1 = \beta\/$.
Hence $ \lambda_s = w_1 (\rho - 2 \beta)\/$, etc.  Thus, the set $
A := \{ \lambda + \rho | \lambda \in F\}\/$ coincides with the set
$
\{ w (\rho - j \beta) | w \in W^\sharp\/$, $ j \in \ZZ_+ \}\/$.  But
by Lemma~2.1, $ w ( e^\rho \R ) = \epsilon (w) e^\rho \R\/$.  This
together with (2.6) completes the proof.\blackbox
\endremarks

\beginremarks{Remark}{1}
Recall that $\epsilon (w) = (-1)^m\/$, where $m\/$ is the number
of reflections with respect to roots from $\Delta_{0+}\/$ entering
in a decomposition of $w\/$ as a product of reflections.  We let
$\check{\varepsilon} (w) = (-1)^{\check{m}}\/$, where $\check{m}\/$
denote the number of reflections occurring in this decomposition
with respect to roots from $\bar\Delta_{0+}\/$.  Then one easily
deduces from (2.4) the formula
$$
e^\rho \check{R} = \Sum_{w \in W^\sharp} \check{\epsilon} (w) w
{e^\rho \over \prod_{\beta \in S} (1 - e^{- \beta})}.
\leqno(2.7)
$$
\endremarks

\beginsection 3 Finite-dimensional irreducible $\gg\/$-modules

Let $\lambda \in \hh^\ast\/$.  The {\it atypicality\/} of
$\lambda\/$, denoted by $ \atp \lambda\/$, is the maximal number of
linearly independent roots $\beta_i\/$ such that $(\beta_i | \beta_j
) = 0\/$ and $( \lambda | \beta_i ) = 0\/$ for all $i\/$ and $j\/$.
Such a set $\{ \beta_i \}\/$ is called a $\lambda\/$-{\it maximal
isotropic\/} subset of $\Delta\/$.  (For example, a maximal
isotropic subset $S \subset \Pi\/$ is a $\rho\/$-maximal isotropic
subset.)

\beginproclaim {Proposition} {1}
Given two $\lambda\/$-maximal isotropic subsets $ \{ \beta_i \}\/$
and $ \{ \beta'_i \}\/$ of $\Delta\/$, there exists $w \in W\/$ such
that $w (\lambda ) = \lambda\/$ and $w ( \beta_i ) = \pm
\beta'_{\sigma (i)}\/$ for some permutation $\sigma\/$.
\endproclaim

Let $V\/$ be a finite-dimensional irreducible $\gg\/$ module.  If we
fix a set of positive roots $\Delta_+\/$, we may talk about the
highest weight $\Lambda\/$ of $V\/$ and about the $\rho\/$.  If
$\Delta'_+\/$ is obtained from $\Delta_+\/$ by a simple $\alpha\/$-
reflection, where $( \alpha | \alpha ) = 0\/$, and if $\Lambda'\/$
is the highest weight for $\Delta'_+\/$, then we clearly have
$$
\Lambda' = \Lambda - \alpha \quad \hbox{if} \quad
( \Lambda | \alpha ) \neq 0; \qquad
\Lambda' = \Lambda \quad \hbox{if} \quad
( \Lambda | \alpha ) = 0.
\leqno (3.1)
$$
Hence by (2.1) we obtain:
$$
\eqalign{%
\Lambda' + \rho' = \Lambda + \rho & \quad \hbox{if} \quad
  ( \Lambda + \rho | \alpha ) \neq 0, \cr
\Lambda' + \rho' = \Lambda + \rho + \alpha & \quad \hbox{if} \quad
  ( \Lambda + \rho | \alpha ) = 0.}
\leqno (3.2)
$$

\beginproclaim {Corollary} {1}
For the $\gg\/$-module $V\/$, $\atp (\Lambda + \rho) \/$ is
independent of the choice of $\Delta_{+}\/$.
\endproclaim

Thus, we may talk about the atypicality of the $\gg\/$-module $V\/$:
$\atp V := \atp (\Lambda + \rho)\/$, which is independent of the
choice of $\Delta_+\/$.  Note that $ \atp V \leq \mydef \gg\/$.

Choose a $\Lambda + \rho\/$-maximal isotropic subset $S_\Lambda\/$
in $\Delta_+\/$ and consider the following rational exponential
function:
$$
\eqalign{
\Sigma_\Lambda & := \sum_{w \in W} \epsilon (w) w
\biggl(
 e^{\Lambda + \rho} \prod_{\beta \in S_\Lambda}
 \left(
   1 + e^{-\beta}
 \right)^{-1}
\biggr), \cr
\check\Sigma_{\Lambda} & := \sum_{w \in W} \check\epsilon (w) w
\biggl(
 e^{\Lambda + \rho} \prod_{\beta \in S_\Lambda}
 \left(
  1 - e^{-\beta}
 \right)^{-1}
\biggr).}
$$
Denote by $j_\Lambda\/$ the coefficient of $e^{\Lambda + \rho}\/$ in
the regular exponential function \break
$\prod_{\alpha \in \bar\Delta_{1+}} ( 1 + e^{-\alpha} )
\sum_\Lambda\/$ and, provided that $j_\Lambda
\neq 0\/$, let
$$
\eqalign{%
{\rm ch}_\Lambda & = j^{-1}_\Lambda e^{-\rho} R^{-1} \Sigma_\Lambda,
    \hbox{ sch}_\Lambda = j^{-1}_\Lambda e^{-\rho} \check{R}^{-1}
    \check\Sigma_\Lambda \cr
{\rm dim}_\Lambda & = {\rm ch}_\Lambda (0), \hbox{ sdim}_\Lambda =
\hbox{sch}_\Lambda (0).}
$$
We shall write sometimes ${\rm ch}_{\Lambda, \Delta_+,
S_\Lambda},\ldots\/$ in place of ${\rm ch}_\Lambda, \ldots\/$ to
emphasize the dependence of $\Delta_+\/$ and $S_\Lambda\/$.

\beginproclaim{Definition}{1}
The $ \gg\/$-module $ V\/$ is called {\it tame\/} with respect to a
subset of positive roots $ \Delta_+\/$ and a subset $ S_\Lambda
\subset \Delta_+\/$, where $ \Lambda\/$ is the highest weight of $
V\/$ with respect to $ \Delta_+\/$ and $ S_\Lambda\/$ is a $ \Lambda
+ \rho\/$-maximal isotropic subset of $ \Delta_+\/$, if $ j_\Lambda
\neq 0\/$ and the character of $ V\/$ is given by
$$
\ch V = \ch_{\Lambda, \Delta_+, S_\Lambda}.
\leqno(3.3)
$$
\endproclaim

\beginremarks {Remarks} {1}
(a) Formula (3.3) implies
$$
\sch V = \pm \sch_{\Lambda, \Delta_+, S_\Lambda}.
$$
Here $+$ or $-$ occurs according as the highest weight vector is
even or odd.

(b)
     If $V\/$ is tame, then $\dim V = \dim_\Lambda, \sdim V = \pm
     \sdim_\Lambda\/$.
\endremarks

\beginremarks{Remarks} {2}
(a)  Suppose that $\Delta'_+\/$ is obtained from $\Delta_+\/$ by a
simple $\alpha\/$-reflection, where $ (\alpha | \alpha) = 0\/$.  If
$ ( \Lambda + \rho | \alpha ) \neq 0\/$, then $\ch_{\Lambda,
\Delta_+, S_\Lambda} = \ch_{\Lambda', \Delta'_+, S_\Lambda}\/$.  If
$ ( \Lambda + \rho | \alpha ) = 0\/$ and $\alpha \in \Pi \bigcap
S_\Lambda\/$, we let $ S'_\Lambda = ( S_\Lambda \setminus \{\alpha\} )
\bigcup \{ - \alpha \}\/$; then $ \ch_{\Lambda, \Delta_+, S_\Lambda}
= \ch_{\Lambda, \Delta'_+, S'_\Lambda}\/$.  This follows from (3.1)
and (3.2).  However, if $ ( \Lambda + \rho | \alpha ) = 0\/$ and
$\alpha \in \Pi \setminus S_\Lambda \/$, $\ch_{\Lambda', \Delta'_+,
S_{\Lambda'}}\/$ may be different from $ \ch_{\Lambda, \Delta_+,
S_\Lambda}\/$.

(b)  Let $ S_\Lambda = \{ \beta_1, \ldots, \beta_s \}\/$ and suppose
$ y \in W\/$ is such that $ y ( \Lambda + \rho ) = \Lambda +
\rho\/$, $ y ( \beta_i ) = \beta_i \/$ for $ i = 2, \ldots, s\/$,
and $ y (\beta_1) = \epsilon (y) \beta'_1\/$, where $ \beta'_1 \in
\Delta_+\/$.  Let $ S'_\Lambda = \{  \beta'_1, \beta_2, \ldots,
\beta_s\}\/$.  Then $\ch_{\Lambda, \Delta_+, S_\Lambda} =
\ch_{\Lambda, \Delta_+, S'_\Lambda}\/$.

(c) Theorem 2.1 states that the 1-dimensional $\gg\/$-module is tame
with respect to $ (\Delta_+, S_0)\/$ if $ \Pi \supset S_0\/$.  Note
that then
$$
j_0 = \left| W / W^\sharp \right|.
\leqno (3.4)
$$
The number $ j_0\/$ is listed below:
$$
\vbox{
\halign{\hfill$#$ & $#$\hfill \cr
A (m, n) (m \geq n ) : (n + 1)!; \; B (m, n) : n! 2^n \hbox{ if } m
> n,\; 2^m m! \hbox{ if } m \leq n;\; C (n) : 1; \cr
D (m, n): n! 2^n\/$ if $m \geq n + 1; 2^{m-1} m!
     \hbox{ if } m < n + 1; \quad D (2, 1; \alpha ),\; F (4),\;
     G (3) : 2. \cr
}% End of halign
}% End of vbox
$$
\endremarks

\beginproclaim{Proposition}{2}
Let $V\/$ be a finite-dimensional irreducible $\gg\/$-module.
Choose a set of positive roots $\Delta_+\/$ and let $\Lambda\/$
denote the highest weight of $V\/$.

(a) ({\rm cf.} [K4]) Let $V \/$ be a typical $\gg\/$-module, {\rm
i.e.}, $( \Lambda + \rho | \alpha ) \neq 0\/$ for $\alpha \in
\bar\Delta_1\/$ (equivalently:  $\atp V = 0\/$).  Then $V\/$ is tame
with respect to $(\Delta_+, \emptyset )\/$.

(b) ({\rm cf.} [BL], [JHKT2, VJ2]) If $\gg\/$ is of type $A (m,
n)\/$ or $C (n)\/$ and $\atp V = 1\/$, then $V\/$ is tame with
respect to $ (\Delta_+, S_\Lambda )\/$ for any choice of
$S_\Lambda\/$ and $j_\Lambda = 1\/$.  (Note that $\atp V = 0\/$ or
$1$ if $\gg\/$ is of type $A (m, 0)\/$ or $C (n)\/$.)

(c) If $( \Lambda + \rho | \alpha ) > 0\/$ for all $\alpha \in
\Delta_{0+}\/$ or $( \Lambda + \rho | \alpha ) \geq 0\/$ for all
$\alpha \in \Delta_{0+}\/$ and $\atp (\Lambda + \rho) = 1\/$ then
$j_\Lambda = 1\/$.
\endproclaim

\beginproof
If $V\/$ is a typical $\gg\/$-module, then formula (3.3) with
$S_\Lambda = \emptyset\/$ and some $j_\Lambda > 0\/$ follows from
the description of singular weights of the Verma module $M ( \Lambda
)\/$ and $W\/$-anti-invariance of $e^\rho \R \ch V\/$ ({\it cf.}
[K4]).  But if $j_\Lambda \neq 1\/$, then the stabilizer of
$\Lambda + \rho\/$ in $W\/$ is non-trivial, and since it is
generated by reflections, we would get that $\ch V = 0\/$, which is
impossible, proving (a).

(c) is clear and (b) follows from [BL, JHKT2, VJ2] (or as above).
\endproof

\beginproclaim{Theorem}{1}
Let $V\/$ be a finite-dimensional irreducible $\gg\/$-module with
highest weight $\Lambda\/$ with respect to a subset of positive
roots $\Delta_+\/$.  If $S_\Lambda \subset \Pi \subset \Delta_+\/$,
then the $\gg\/$-module $V\/$ is tame with respect to $( \Delta_+,
S_\Lambda )\/$.

Note that Theorem 2.1 follows from Theorem 3.1 and Remark 3.2b.
\endproclaim

\beginexample{1}
If $\atp V = 0\/$, then $S_\Lambda = \emptyset\/$ and Theorem~3.1
gives the well-known character formula of a typical $\gg\/$-module
({\it cf.} Proposition~3.2a).
\endexample

\beginexample{2}
If $\atp V = 1\/$ and $\gg\/$ is of the first kind, {\it i.e.,\/} of
type $A (m, n)\/$ or $C (n)\/$, then one can always choose
$\Delta_+\/$ such that $S_\Lambda \subset \Pi\/$, hence by
Theorem~3.1 the $\gg\/$-module $V\/$ is tame ({\it cf.}
Proposition~3.2b).  There exist, however, modules of atypicality 3
over $A (4, 3)\/$ that are not tame for any choice of $\Delta_+\/$
and $S_\Lambda\/$ (examples were provided by J.~Van der~Jeugt).
\endexample

\beginexample{3}
 Let now $\gg\/$ be of the second kind, {\it i.e.,\/} of type $B (m,
     n)\/$, $D(m, n)\/$, $D (2, 1; \alpha)\/$, $G(3)\/$ or
     $F(4)\/$.  We exclude $B (0, n)\/$ for which all $V\/$ are
     typical.  Let
     $\Delta_+\/$ be a standard set of positive roots.  Here are the
     corresponding Dynkin diagrams [K2, p.~56]:
$$
\vbox{\halign{$#\hfil$ & \quad $#$ \hfil \cr
B (m, n), m > 0 & \bigcirc - \bigcirc - \ldots -
\otimes_2 - \bigcirc_2 -
\ldots - \bigcirc_2 \Rightarrow
  \bigcirc_2 \cr
B (0, n) & \bigcirc - \bigcirc - \ldots - \bigcirc - \bigcirc
\Rightarrow \bigcirc \cr
D (m, n), m \geq 2 & \bigcirc - \bigcirc - \ldots -
  \otimes_2 - \bigcirc_2
  - \ldots - \bigcirc_2
  \hskip-.5ex\matrix{
    & \bigcirc_1 \cr
    \cr
    & \bigcirc_1 \cr
  } \cr
D (2, 1; \alpha ) & {}^2\!\otimes
  \hskip-.5ex\matrix{
    & \bigcirc_1 & \cr
    & & \cr
    & \bigcirc_1 & \cr
  } \cr
G (3) & \otimes_2 -
\bigcirc_4 \Llleftarrow
  \bigcirc_2 \cr
F (4) & \otimes_2 -
\bigcirc_3 \Leftarrow
  \bigcirc_2 - \bigcirc_1 \cr
}}
$$
Let \( \tilde\theta\/\) be the only simple root of \(
\Delta_{0+}\/\) which is not in \( \Pi\/\); its coefficients in
terms of \( \Pi\/\) are given above (note that \( \tilde\theta =
\theta\/\) in all cases except for \( B (m, n)\/\) and \( D(m,
n)\/\) with \( n > 1\/\)).  Let \( b = - ( \rho |
\tilde\theta^\vee)\/\) (as usual, \( \alpha^\vee := 2 \alpha /
(\alpha | \alpha)\/\) for \( \alpha \in \Delta_0\/\)); the values of
\( b\/\) are as follows:
$$
\matrix{
B (m, n): m - {1 \over 2}, & D (m, n): m - 1, & D (2, 1: \alpha): 1,
& G ( 3 ): {5 \over 2}, & F(4) : 3.
}
$$
Let \( V\/\) be a finite-dimensional irreducible \( \gg\/\)-module
and let \( \Lambda\/\) be its highest weight.  Let \( k = ( \Lambda
| \tilde\theta^\vee)\/\).  Recall that \( \dim V < \infty\/\)
implies that \( k_i := (\Lambda | \alpha^\vee_i ) \in \ZZ_+\/\) for
all \( \alpha_i \in \Pi \bigcap \Delta_0\/\) and that \( k \in
\ZZ_+\/\).  For \( k > b\/\) these conditions are sufficient, and
for \( k \leq b\/\) there are some extra conditions listed in
[K2, p.~84] (note that \( b\/\) there slightly differs from \(
b\/\) here).  Suppose that \( \atp V = 1\/\) (this
is always the case for a non-typical \( V\/\) over \( \gg\/\) of
type \( B (m, 1)\/\), \( B (1, n)\/\), \( D (m, 1)\/\), \( D (2, 1;
\alpha )\/\), \( G (3)\/\), and \( F (4)\/\)), and let \( S_\Lambda
= \{ \beta \}\/\) (the choice of \( \beta\/\) may not be unique).
We have:  \( j_\Lambda = 1\/\) if \( k \geq b\/\) and \( j_\Lambda
= 2\/\) if \( k < b\/\).  If the \( \gg\/\)-module with the highest
weight \( r_{\tilde\theta} ( \Lambda + \rho ) - \rho\/\) is not
finite-dimensional (which is always the case when \( k \leq b\/\) or
\( k > 2b\/\)), then \( V\/\) is tame, {\it i.e.,\/}
$$
\ch V = \ch_\Lambda.
$$
Otherwise we have:
$$
\ch V = \ch_\Lambda - \ch_{r_{\tilde\theta} ( \Lambda + \rho ) -
\rho}.
$$
Furthermore, provided that the highest weight vector is even, we
have:
$$
\sign ( \sdim_\Lambda ) = ( -1)^{\ell (w)} ,
$$
where \( w \in W\/\) is the shortest element such that \( w ( \beta
) \in \Pi\/\).  Here is a complete list of non-tame modules over
exceptional Lie superalgebras (we write \( \Lambda = ( k ; k_2,
k_3, \ldots )\/\)):
$$
\vbox{\halign{$#\hfil$ & \quad $#$ \hfil \cr
D (2, 1; \alpha ) : & \Lambda = \theta = (2; 0, 0) : \ch V =
     \ch_\theta - 1 . \cr
G (3): & \Lambda = (3; 0, n) : \ch V = \ch_{(3; 0, n)} - \ch_{(2; 0,
   n)}. \cr
     & \Lambda = (5; 0, 0) :  \ch V = \ch_{(5; 0, 0)} - 1. \cr
F (4): & \Lambda = ( 4; 0, n, 0 ) : \ch V = \ch_{(4; 0, n, 0)} -
 \ch_{(2; 0, n, 0)} . \cr
     & \Lambda = (6; 0, 0, 0) :  \ch V = \ch_{(6; 0, 0, 0)} - 1 .
     \cr
}}
$$
In the case of $ D (2, 1; \alpha )\/$ this result is consistent with
[VJ1].
\endexample

A proof of these results will be given elsewhere.  Here we note that
the same proof as of Theorem~2.1 shows that for $ \gg\/$ of defect
1 and atypical $ V\/$ with the highest weight $ \Lambda\/$ such that
$ ( \Lambda + \rho | \beta ) = 0\/$, $ \beta \in \bar\Delta_{1+}\/$
and such that $ k < b\/$ for $ \gg\/$ of the second kind one has:
$$
e^\rho \R \ch V =
\sum_{w \in W^\sharp} \epsilon (w) w
\left(
 e^{\Lambda + \rho} / (1 + e^{-\beta})
\right) .
\leqno(3.5)
$$

Given integers \( a\/\) and \( b\/\) such that \( -1 \leq a \leq
b\/\), we let
$$
C (a, b) = \sum^b_{j=1} 2^{b-j} {j \choose a} \hbox{ if } a \geq 0;
     \quad C ( -1, b ) = \delta_{-1,b} 2^{b+1}.
\leqno (3.6)
$$
Let \( N_\epsilon = | \Delta_{\epsilon +} |\/\), \( \epsilon = 0\/\)
or 1.

\beginproclaim{Theorem}{2}
Let \( \Lambda \in \hh^\ast\/\), and let \( S_\Lambda = \{ \beta_1,
\ldots, \beta_s\}\/\).  Then (here and further \( | \vec k |\/\)
stands for \( \sum_i k_i\/\)):
$$
\dim_\Lambda = {2^{N_1 - N_0 - s} \over {j_\Lambda}}
     \sum_{\vec k \in \ZZ^s_+ \atop |\vec k| \leq N_0} (-1)^{|
     \vec k|} C ( | \vec k| + s - 1, N_0 + s - 1 )
     \prod_{\alpha \in \Delta_{0+}}
     { (\Lambda + \rho - \sum_i k_i \beta_i | \alpha ) \over (
     \rho_0 | \alpha )}.
$$
\endproclaim

The proof of the theorem is based on the following simple lemma.

\beginproclaim{Lemma}{1}
Let \( P (x_1, \ldots, x_s )\/\) be a polynomial of degree \( \leq
d\/\) in the indeterminates \( x_1, \ldots, x_s\/\).

(a) In the domain \( |x_i| < 1\/\) one has
\[
\sum_{\vec n \in \ZZ^s_+} (-1)^{|\vec n|} \vecx^{\vecn} P(\vecn) =
\sum_{\vecn \in \ZZ^s_+ \atop |\vecn| \leq d}\;
\sum_{\veck \in \ZZ^s_+ \atop k_i \leq n_i} (-1)^{|\veck|}
\pmatrix{\vecn \cr \veck} P (\veck)
\prod^s_{j=1} {x^{n_j}_j \over (1 + x_j)^{n_j +1}}.
\leqno(3.7)
\]
Here and further we use the usual notation:
\[
\vecx^{\vecn} = x^{n_1}_1 \ldots x^{n_s}_s, \quad
\pmatrix{\vecn \cr \veck} = \pmatrix{n_1 \cr k_1}
\ldots \pmatrix{n_s \cr k_s}.
\]
(b) The value of the rational function on the right-hand
side of (3.7) at \( x_1 = \ldots = x_s = 1\/\) equals
\[
2^{-s-d} \sum_{\vecn \in \ZZ^s_+ \atop |\vecn| \leq d}
(-1)^{|\vecn|} C
(|\vecn| + s -1, d + s -1) P(\vecn).
\leqno(3.8)
\]
\endproclaim

\medbreak\noindent {\it Proof. \enspace}
It suffices to prove (3.7) for a polynomial \( P(x) = \pmatrix{x \cr
j}\/\), \( 0 \leq j \leq d\/\).  But this follows from the
well-known formulas:
\[
\eqalign{
\sum^n_{k=0} (-1)^k \pmatrix{n \cr k} \pmatrix{k \cr j} & =
  (-1)^n \delta_{n,j}, \cr
\sum_{n\in \ZZ_+} (-x)^n \pmatrix{n \cr j} & =
  {(-x)^j \over (1 + x)^{j+1}} \quad \hbox{in the domain}\quad |x|
  < 1. \cr
}% End of eqalign
\]

(b) follows from (a) and the formula
$$
\sum_{\vecn \in \ZZ^s_+ \atop |\vecn| = N}
\pmatrix{\vecn \cr \veck} =
\pmatrix{N + s - 1 \cr |\veck| + s - 1} .
\eqno\blackbox
$$

\beginremarks{Remark}{3}
It is clear from the proof that one can take the summation over \(
\vecn\/\) in the right-hand side of (3.7) and in (3.8) over the set
\( [P] \in \ZZ^s_+\/\) defined as follows.  We let \( [
\vecx^{\veca} ] = \{ \vecb \in \ZZ^s_+ | b_i \leq a_i \}\/\) and let
\( [P]\/\) be the union of \( [ \vecx^{\veca} ]\/\) over all the
monomials occuring in \( P\/\) with a non-zero coefficient.
\endremarks

\medbreak\noindent{\bf Proof of Theorem 3.2.} \enspace
Given \( \lambda \in \hh^\ast\/\), define the specialization \( F_{t
\lambda}\/\) by
\[
F_{t \lambda} (e^\mu) = e^{t (\lambda | \mu)}, \quad \mu \in
\hh^\ast.
\]
First, we compute the following specialization, where \( | x_i | <
1\/\) for all \( i\/\) and \break
\( R_0 = \prod_{\alpha \in \Delta_{0+}} (1 - e^{-\alpha} )\/\):
\[
\displaylines{
\qquad F_{t \rho_0}
\left(
   {1 \over e^{\rho_0} R_0} \sum_{w \in W} \epsilon (w) w
   {e^{\Lambda + \rho} \over \prod_i ( 1 + x_i e^{-\beta_i})}
\right) \hfill \cr
\vbox{\halign{\hfil$\displaystyle #$\hfil & $\displaystyle #$\hfill
 \cr
 = &
  \sum_{\vec n \in \Bbb Z^s_+} (-1)^{|\vec n|} \vec x^{\vec n}
  {\displaystyle
\sum_{w \in W} \epsilon (w) e^{t ( w ( \Lambda + \rho - n_1 \beta_1
 - \ldots - n_s \beta_s) | \rho_0)}
 \over F_{t \rho_0} (e^{\rho_0} R_0 )} \cr
 = &
   \sum_{\vec n \in \Bbb Z^s_+} (-1)^{|\vec n|} \vec x^{\vec n}
   { F_{t ( \Lambda + \rho - \Sigma n_i \beta_i)} (e^{\rho_0} R_0)
   \over F_{t \rho_0} (e^{\rho_0} R_0 )}. \cr
} % End of halign
} % End of vbox
} % End of displaylines
\]
Hence we have in the domain \( |x_i| < 1\/\):
\[
\eqalign{
& \displaystyle
  \lim_{t \rightarrow 0} F_{t \rho_0}
  \left(
    {1 \over e^{\rho_0} R_0} \sum_{w \in W} \epsilon (w) w
    {e^{\Lambda + \rho} \over \prod_i (1 + x_i e^{-\beta_i})}
  \right) \qquad\qquad \cr
& {} = \displaystyle
  \sum_{\vec n \in \ZZ^s_+} (-1)^{|\vec n|} \vec x^{\vec n}
  \prod_{\alpha \in \Delta_{0+}}
  {\displaystyle
   (\Lambda + \rho - \sum_i n_i \beta_i | \alpha ) \over ( \rho_0 |
    \alpha )}. \cr
} % End of eqalign
\leqno(3.9)
\]
Note that the expression under the limit at the left-hand side of
(3.9) is a continuous function in the domain \( |t| < \epsilon\/\)
and \( |x_i - 1| < \epsilon\/\) for sufficiently small \(
\epsilon\/\).  Hence we may calculate
\[
\lim_{t \rightarrow 0} F_{t \rho_0}
\left(
     {1 \over e^{\rho_0} R_0} \sum_{w\in W} \epsilon (w) w
     {e^{\Lambda + \rho} \over \prod_i (1 + e^{-\beta_i})}
\right)
\]
by applying Lemma 3.1a to the right-hand side of (3.9) and then
evaluating it at \( x_1 = \ldots = x_s = 1\/\) using Lemma 3.2b.
Since \( \dim_\Lambda = \lim_{t \rightarrow 0} F_{t \rho_0}
\ch_\Lambda\/\), we obtain the result. \blackbox\par\medbreak

\beginproclaim{Theorem}{3}
Let \( \Lambda \in \hh^\ast\/\), let \( S_\Lambda\/\) be a \(
\Lambda + \rho\/\)-isotropic subset and let \( s = | S_\Lambda
|\/\).  Consider the root system \( M_\Lambda := \{ \alpha \in
\bar\Delta_0 | \alpha \perp \SLambda\}\/\), let \(
\rho^0_\Lambda\/\) be the half of the sum of roots from \(
M^+_\Lambda := M_\Lambda \bigcap \Delta_+\/\) and let \( L_\Lambda
(\lambda)\/\) denote the irreducible module with highest weight \(
\lambda\/\) over the reductive Lie algebra with root system \(
M_\Lambda\/\).  Let \( a = n - m - 1\/\) for \( \gg\/\) of type \(
B(m,n)\/\) with \( m < n\/\) and let \( a = 0\/\) in all other
cases.  Then
\[
\leqalignno{
\left|
  \sdim_\Lambda
\right| & = {| W / W^\sharp|}{2^{-a} j^{-1}_\Lambda}
\dim L_\Lambda (\Lambda + \rho - \rho^0_\Lambda) \hbox{ if } s =
\mydef \gg, & ({\rm 3.10a}) \cr
\sdim_\Lambda & = 0 \hbox{ otherwise.} & ({\rm 3.10b}) \cr
}% End of leqalignno
\]
\endproclaim

\beginproof
Let \( W^\Lambda\/\) be the subgroup of \( W\/\) generated by
reflections with respect to the roots from \( M_\Lambda\/\) and let
\( W_1\/\) be a set of left coset representatives, so that \( W =
W_1 W^\Lambda\/\).  Due to (2.3), formula (3.4) can be rewritten
using the Weyl character formula for the root system \(
M_\Lambda\/\) as follows:
\[
\displaylines{% Beginning of displaylines
\quad j_\Lambda \prod_{\alpha \in \Delta_{1+} \setminus
\bar\Delta_{1+}} ( 1 + e^{-\alpha} ) \prod_{\alpha \in
\bar\Delta_{0+}} (1 - e^{-\alpha} ) e^\rho \sch_\Lambda \hfill
\cr
\hfill
= \prod_{\alpha \in \bar\Delta_{1+}} ( 1 - e^{-\alpha} )
\sum_{w \in W_1} \check\epsilon (w)
{
  \prod_{\alpha \in M^+_\Lambda} (1 - e^{- w (\alpha)}) \over
  \prod_{\alpha \in \SLambda} (1 - e^{-w (\alpha)} )
}
w \left(
  e^{\rho^0_\Lambda} \ch L_0 ( \Lambda + \rho - \rho^0_\Lambda )
\right).\quad \cr
}% End of displaylines
\]
For \( \lambda \in \hh^\ast\/\) such that \( ( \lambda | \alpha )
\neq 0\/\) for all \( \alpha \in \Delta\/\), we apply to both sides
of this equality the specialization \( F_{t \lambda}\/\) to obtain
as \( t \rightarrow 0\/\):
\[
\displaylines{% Beginning of displaylines
\quad j_\Lambda 2^{| \Delta_{1+} \setminus \bar\Delta_{1+} |}
\prod_{\alpha \in \bar\Delta_{0+} \setminus
\bar\Delta_{1+}} ( \lambda | \alpha ) \sdim_\Lambda = \hfill
\cr
\hfill{} t^{| \bar\Delta_{1+}| + |M^+_\Lambda| -
|\bar\Delta_{0+}| - |\SLambda|}
\left(
  \sum_{w \in W_1} \check\epsilon (w) \prod_{\alpha \in M^+_\Lambda
  \setminus \SLambda}
  \left(
    \lambda | w (\alpha )
  \right)
\right) \dim L_\Lambda ( \Lambda + \rho - \rho^0_\Lambda ) +
o (t).\cr
}% End of displaylines
\]
Due to Lemma 2.2d, if \( s < \mydef \gg\/\), the right-hand side
tends to 0 as \( t \rightarrow 0\/\), proving (3.10b), and if \( s
= \mydef \gg\/\), we obtain as \( t \rightarrow 0\/\):
$$
\sdim_\Lambda = j^{-1}_\Lambda c (\lambda, \Delta_+,
S_\Lambda ) \dim L_\Lambda ( \Lambda + \rho - \rho^0_\Lambda ),
\leqno(3.11)
$$
where
$$
 c (\lambda, \Delta_+, S_\Lambda ) =
     \prod_{\alpha \in \bar\Delta_{0+} \setminus
          \bar\Delta_{1+}} ( \lambda | \alpha )
          \sum_{w \in W_1} \check\epsilon (w) \prod_{\alpha \in
          M^+_\Lambda \setminus S_\Lambda} (\lambda | w (\alpha)
          )
\leqno(3.12)
$$
is a constant which, in fact, is independent of the choice of \(
\lambda\/\) (since all other factors in (3.12) are independent of \(
\lambda\/\)).  It follows that the constant \( c = c (\lambda,
\Delta_+, \SLambda )\/\) is independent up to a sign of the choice
of \( (\Delta_+, \SLambda )\/\) as well.  Recall now that, according
to Theorem 2.1, the 1-dimensional module is tame with respect to a
certain pair \( (\Delta'_+, S')\/\).  Applying (3.11) to the
1-dimensional \( \gg\/\)-module (for which \( \sdim = \pm 1\/\)) and
using (3.4), we obtain:
\[
1 = \pm | W /W^\sharp |^{-1} c \dim L_0 (\rho' - {\rho^0_0}' ).
\]
Finally a case by case inspection shows that \( \dim L_0 (\rho' -
{\rho^0_0}' ) = 2^a\/\), proving (3.10a).

In the case $ \gg = C (n)\/$ a formula equivalent to (3.10b) was
obtained in [VJ2].
\endproof

\beginconjecture{1}
$ \sdim V = 0\/$ if and only if $ \atp V < \mydef \gg\/$.
\endconjecture

\beginexample{4}
(a) if $ V\/$ is a typical $ \gg\/$-module then formulas (3.6) and
(3.10) turn into the known formulas [K4]:
$$
\eqalign{
\dim V & = 2^{N_1} \prod_{\alpha \in \Delta_{0+}}
     {(\Lambda + \rho | \alpha) \over (\rho_0 | \alpha)}, \cr
\sdim V & = 0 \hbox{ if } \gg \neq B (0, n),\;
   = \prod_{\alpha \in \bar\Delta_{0+}} (\Lambda + \rho |
          \alpha ) / (\rho | \alpha ) \hbox{ if } \gg = B (0, n).
          \cr
}% End of eqalign
$$

(b) Let $ \gg\/$ be an exceptional Lie superalgebra and let $ V\/$
be a finite-dimensional irreducible non-typical $ \frak{g}\/$-module
with highest weight $ \Lambda\/$, so that $ \atp V = 1\/$.  We
keep the notation of Example 3.3.

$\gg = D(2, 1; \alpha) : (\alpha_1 | \alpha_1) = 0\/$,
$(\alpha_1 | \alpha_2 ) = -1\/$,
$(\alpha_1 | \alpha_3 ) = -\alpha\/$,
$(\alpha_2 | \alpha_2 ) = 2\/$,
$(\alpha_3 | \alpha_3 ) = 2 \alpha\/$,
$(\alpha_2 | \alpha_3 ) = 0\/$.  Assume that $ V\/$ is neither the
1-dimensional nor the adjoint module.  Then $ \sdim V = \pm 2\/$ and
$ {1 \over 2} \dim V\/$ is given by the following formulas, where $
m = k_2\/$, $ n = k_3\/$:
$$
\eqalign{
4 kmn + 2k (m + n) - 2n + 1 & \hbox{ if } \beta = \alpha_2, \cr
4 kmn + 2 (km + 3kn - mn) + 4 (k - n) -1 & \hbox{ if } \beta =
   \alpha_1 + \alpha_2, \cr
4 kmn + 2 (3km + kn - mn) + 4 (k - m) - 1 & \hbox{ if } \beta =
   \alpha_2 + \alpha_3, \cr
4 kmn + 2 (3km + 3kn - mn) + 4 (2k - m - n) - 7 & \hbox{ if } \beta
   = \alpha_1 + \alpha_2 + \alpha_3.
}% End of eqalign
$$
(Recall that for a typical $ V\/$ one has [K4]: $ \dim V = 16 (m
+ 1) (n + 1) (k - 1)\/$ and $ \sdim V = 0\/$.)

$ \gg = G(3)\/$ ($m = k_2\/$, $ n = k_3\/$):

\noindent
If $ \beta = \alpha_1\/$, then $ \dim V = 1\/$; $ \beta \neq
\alpha_1 + \alpha_2\/$.  If $ \beta = \alpha_1 + \alpha_2 +
\alpha_3\/$, then $ k = 2\/$, $ m = 0\/$ and $ \sdim V = 2n + 3\/$.
If $ \beta = \alpha_1 + 3 \alpha_2 + \alpha_3\/$, then $ 2k =
m+6\/$; for $ k = 3\/$ we have $ \sdim V = -2n - 3\/$, for $ k >
3\/$ we have $ \sdim V = -2 (m + 2n + 3)\/$.  If $ \beta = \alpha_1
+ 3\alpha_2 + 2\alpha_3\/$, then $ 2k = m + 3n + 9\/$ and $ \sdim V
= 2 (m + n + 2)\/$.  If $ \beta = \alpha_1 + 4 \alpha_2 +
2\alpha_3\/$, then $ 2k = 2m + 3n + 10\/$; for $ k = 5\/$ we have $
\dim V = 321\/$, $ \sdim V = -1\/$, for $k > 5\/$ we have $ \sdim V
= -2 (n + 1)\/$.

$ \gg = F(4)\/$ $(l = k_2\/$, $ m = k_3\/$, $ n = k_4\/$):

\noindent
If $ \beta = \alpha_1\/$, then $ \dim V = 1\/$; $ \beta \neq
\alpha_1 + \alpha_2\/$.  If $ \beta = \alpha_1 + \alpha_2 +
\alpha_3\/$, then $ k = 2\/$, $ l = n = 0\/$ and $ \sdim V = (m +
2)^3\/$.  If $ \beta = \alpha_1 + 2 \alpha_2 + \alpha_3\/$, then $
3k = l - 2n + 8\/$; if $ \beta = \alpha_1 + \alpha_2 + \alpha_3 +
\alpha_4\/$, then $ 3 k = 2n - l + 10\/$; in both cases $ k \geq
3\/$ and $ \sdim V = - (m + 1) ( l + m + n + 3) (l + 2m + n + 4)\/$.
If $ \beta = \alpha_1 + 2 \alpha_2 + \alpha_3 + \alpha_4\/$, then $
3 k = l + 2n + 12\/$; if $ k = 4\/$, then $ l = n = 0\/$ and $ \sdim
V = (m + 2)^3\/$; if $ k > 4\/$, then $ \sdim V = (l + m + 2) (m +
n + 2) (l + 2m + n + 4)\/$.  If $ \beta = \alpha_1 + 2 \alpha_2 + 2
\alpha_3 + \alpha_4\/$, then $ 3k = l + 4m + 2n + 16\/$ and $ \sdim
V = - (l + m + 2) (n + 1) (l + m + n + 3)\/$.  If $ \beta = \alpha_1
+ 3\alpha_2 + 2\alpha_3 + \alpha_4\/$, then $ 3k = 3l + 4m + 2n +
18\/$; if $ k = 6\/$, then $ l = m = n = 0\/$ and $ \sdim V = 1\/$;
if $ k > 6\/$, then $ \sdim V = (m + 1) (n + 1) (m + n + 2)\/$.
\endexample

\sectnum=4
\num=0
\par{\vskip0pt plus .2\vsize\penalty-250
     \vskip0pt plus-.2\vsize\bigskip\vskip\parskip
%     \message{4 The denominator identity for
%                affine superalgebras and its specializations}
\vbox{
  \halign{#\hfill \cr
    \indent\indent
    \S {\bf 4.\enspace The denominator identity
      for affine superalgebras} \cr
      \hphantom{\indent\indent\S {\bf 4.\enspace}}
      {\bf and its specializations.}\cr
}% End of halign
}% End of vbox
\nobreak\smallskip\indent}
Let $ \gg\/$, as before, be a finite-dimensional simple Lie
superalgebra over $ \CC\/$ of rank $ \ell\/$ with a non-degenerate
even invariant
bilinear form $(.|.)$ (see Section 1).  Let $ \hat\gg = \CC [t, t^{-
1}] \otimes \gg \oplus \CC C \oplus \CC d\/$ be the associated to $
\gg\/$ affine superalgebra.  The $ \ZZ_2\/$-gradation of $
\hat\gg\/$ extends from that of $ \gg\/$ by letting $ \deg t = 0\/$,
$ \deg C = \deg d = 0\/$.  Denoting $ a (n) = t^n \otimes a\/$ for
$ a \in \gg\/$, $ n \in \ZZ\/$, we have the following commutation
relations ($ a, b \in \gg\/$; $ m, n \in \ZZ\/$):
$$
\eqalign{
[ a(m), b(n) ] &= [a,b] (m+n) + m \delta_{m, -n} (a | b) C, \cr
[ d, a(m) ] &= m a(m), \quad [C, \hat\gg ] = 0.
}% End of eqalign
$$
We identify $ \gg\/$ with the subalgebra $ 1 \otimes \gg\/$ of $
\hat \gg\/$.  The bilinear form $(.|.)$ extends from $ \gg\/$ to an
even invariant bilinear form on $ \hat\gg\/$ by:
$$
\displaylines{
\left( a(m) | b(n)\right) = \delta_{m, -n} (a | b), \qquad
     \left( \CC [t, t^{-1}] \otimes \gg | \CC C + \CC d\right) = 0,
     \cr
( C | C ) = ( d | d ) = 0, \qquad (C | d) = 1.
}% End of displaylines
$$
Let $ \hat\hh = \hh + \CC C + \CC d\/$ be the Cartan subalgebra
of $ \hat\gg\/$ and let $ \hat \nn_+ = \nn_+ \oplus \left(
\bigoplus_{j > 0} t^j \otimes \hh \right)\/$.  We identify $
\hat\hh^\ast\/$ with $ \hat\hh\/$ using the bilinear form $(.|.)$.

Given a subset $ S \subset \hh\/$, we let $ \tilde S = \{ \alpha
+ n C | \alpha \in S, n \in \ZZ \} \setminus \{ 0 \}\/$.  Then the
set of roots of $ \hat\gg\/$ with respect to $ \hat\hh\/$ is $
\hat\Delta = \tilde\Delta_0 \cup \{0\}\tilde{}\/.$, the set of even
(resp.\ odd) roots being $ \hat\Delta_0 = \tilde\Delta_0 \cup \{
0 \}\tilde{}\/$ (resp.\ $ \hat\Delta_1 = \tilde\Delta_1\/$).  The set
of positive roots of $ \hat\Delta\/$ associated to the choice of $
\Delta_+ \subset \Delta\/$ is $ \hat\Delta_+ = \Delta_+ \cup \{
\alpha + n C | \alpha \in \Delta \cup \{ 0 \},\, n > 0 \}\/$, the
set of simple roots being $ \hat\Pi = \{ \alpha_0 := C - \theta \}
\cup \Pi\/$.  As usual the subscript + means the intersection with
$ \hat\Delta_+\/$, {\it e.g.,} $ \tilde\Delta_+ = \tilde\Delta \cap
\hat\Delta_+\/$.

Define $ \hatrho \in \hathh^\ast\/$ by $ \hatrho = \rho + h^\vee
d\/$.  One has:
$$
(\hatrho | \alpha_i ) = {\textstyle {1 \over 2}} (\alpha_i |
\alpha_i ) \hbox{ for } \alpha_i \in \hat\Pi,\quad (\hatrho | d) =
0,\quad (\hatrho | C ) = h^\vee.
$$
As before, for $ \alpha \in \tilde\Delta_0\/$ we let $ \alpha^\vee
= 2 \alpha / ( \alpha | \alpha )\/$.  For $ \alpha \in \hh^\ast\/$
define $ t_\alpha \in \hbox{Aut}\, \hat\hh^\ast\/$ by
$$
t_\alpha ( \lambda ) = \lambda + \lambda (C) \alpha -
\left(
   (\lambda | \alpha) + {\textstyle {1 \over 2}} (\alpha | \alpha)
   \lambda (C)
\right) C.
$$
For an additive subgroup $ S \subset \hh^\ast\/$ the set $ t_S := \{
t_\alpha | \alpha \in S \}\/$ is a subgroup of $ \Aut
\hat\hh^\ast\/$.  Let $ M (\hbox{resp} M^\sharp)\/$ be the $ \ZZ\/$-span of
the $ \alpha^\vee\/$ such that $ \alpha \in \Delta_0\/$ (resp.\ $
\alpha \in \Delta^\sharp_0\/$).  The Weyl group of $ \hat\gg\/$ is
$ \widehat W = W \ltimes  t_M\/$.  The following subgroup of $
\widehat W\/$ is more important, however:  $ \widehat W^\sharp = W
\ltimes t_{M^\sharp}\/$.

Denote by $ \widehat{\Cal F}\/$ the field of meromorphic functions
in the domain $ Y = \{ h \in \hat\hh | $ \allowbreak $\Re (C | h) >
0 \}\/$.  Since $ C\/$ is $ \widehat W\/$-fixed, the group $
\widehat W\/$ acts on $ \widehat{\Cal F}\/$.  Let $ q = e^{-C}\/$;
note that $ |q| < 1\/$ in $ Y\/$.  Given $ \lambda \in \hh^\ast\/$,
we extend it to a linear function $ \tilde\lambda \in
\hat\hh^\ast\/$ by letting $ \tilde\lambda (C) = \tilde\lambda (d)
= 0\/$; this gives us an embedding of the field $ \Cal F\/$ of
rational exponential functions on $ \hh\/$ in the field
$ \widehat{\Cal F}\/$.

Let R be the Weyl denominator of $ \gg\/$ (see Section~2).
Introduce the {\it affine denominator\/} by
$$
\widehat\R = \R \prod^\infty_{n=1} (1 - q^n)^\ell
\prod_{\alpha \in \Delta_0} (1 - q^n e^\alpha)
\prod_{\alpha \in \Delta_1} (1 + q^n e^\alpha )^{-1}.
$$
It is clear that $ \widehat\R \in \widehat{\Cal F}\/$.  As in Section
2, it is easy to check that
$$
w ( e^{\hat\rho} \widehat\R ) = \epsilon (w) e^{\hat\rho}
\widehat\R,\quad
w \in \widehat W.
\leqno(4.1)
$$
Here $ \epsilon : \widehat W \rightarrow \pm 1\/$ is a homomorphism
which extends the homomorphism $ \epsilon : W \rightarrow \pm 1\/$
by $ \epsilon (t_M) = 1\/$.

\beginproclaim{Theorem}{1}
Suppose that $ h^\vee \not= 0\/$.  Then
$$
e^{\hat\rho} \widehat\R = \sum_{\alpha \in M^\sharp} t_\alpha
(e^{\hat\rho} \R).
\leqno(4.2)
$$
\endproclaim

The proof of Theorem 4.1 is based on the following

\beginproclaim{Lemma}{1}
Suppose that $ \Lambda \in \hat\hh^\ast\/$ is such that $ (\Lambda
+ \hat\rho ) (C) \not= 0\/$, and suppose that $ \hat L (\lambda)\/$,
$ \lambda \in \hat\hh^\ast\/$, is an irreducible subquotient of the
Verma $ \hat\gg\/$-module $ \widehat\M (\Lambda)\/$.  Then there
exists a sequence $ \lambda_0 = \Lambda, \lambda_1, \ldots,
\lambda_k = \lambda\/$ such that $ \lambda_i\/$ is obtained from $
\lambda_{i - 1}\/$ ($ 0 < i \leq k\/$) in one of the following two
ways:
\itemitem{(i)}
     $ \lambda_i = \lambda_{i-1} - n \beta_i\/$, where $ \beta_i
     \in \tilde\Delta_{0+}\/$ and $ (\lambda_{i-1} + \hat\rho |
     \beta^\vee_i ) = n \in \NN\/$,
\itemitem{(ii)}
     $ \lambda_i = \lambda_{i-1} - \beta_i\/$, where $ \beta_i \in
   \tilde\Delta_{1+}\/$, $ (\beta_i | \beta_i) = 0\/$, and $
   (\lambda_{i-1} + \hat\rho | \beta_i ) = 0\/$.
\endproclaim

\beginproof
See [K6, Theorem 1b].  (Note that the usual proof
is based on the determinant formula given by [K6, Theorem 1a];
the exponents in this formula should be corrected as in [K8]).
\endproof

A detailed proof of Theorem 4.1 will be given elsewhere.  In the
case $ \mydef \gg = 1\/$ the same proof works as that in Section 2.

\beginproclaim{Theorem}{2}
Let $ \gg\/$ be a simple finite-dimensional Lie superalgebra with a
non-degenerate Killing form, {\rm i.e.,}  $
h^\vee \neq 0\/$.  Choose a set of positive roots $ \Delta_+\/$
such that the set of simple roots contains a maximal set of pairwise
orthogonal isotropic roots $ \{ \beta_1, \ldots, \beta_s\} ( s =
\mydef \gg )\/$.  Let $ \beta = \sum_{i=1}^s \beta_i\/$ and let $
d_\epsilon = \dim \gg_\epsilon\/$, $ \epsilon = 0\/$ or 1.  Then
$$
\leqalignno{
\prod^\infty_{j = 1}
{(1 - q^j)^{d_0}
\over
(1 + q^j)^{d_1}} & =
     {2^{N_1}
     \over | W / W^\sharp |}
     \sum_{\alpha \in M^\sharp}
     q^{{1\over2} h^\vee (\alpha | \alpha) + (\rho + \beta |
          \alpha)}
     \sum_{\vec n \in \ZZ^s_+ \atop |\vec n| \leq N_0}
     {A_\alpha (\vec n)\over\prod_i ( 1 + q^{(\alpha |
     \beta_i)})^{n_i + 1}}, & (4.3) \cr
\noalign{\hbox{where}}
A_\alpha (\vec n) & =
     \sum_{\vec k \in \ZZ^s_+ \atop k_i \leq n_i} (-1)^{|\vec k|}
          \left(
            {\vec n \atop \vec k}
          \right) \prod_{\gamma \in \Delta_{0+}}
          {(\rho + h^\vee \alpha - k_1 \beta_1 - \cdots - k_s
          \beta_s | \gamma)
          \over( \rho_0 | \gamma)}. & (4.4) \cr
}% End of \leqalignno
$$
\endproclaim

\beginproof
{}From Theorems 2.1 and 4.1, and Remark 3.2c we have:
$$
{| W / W^\sharp| \widehat\R \over e^{\rho_0} \R_0} =
\lim_{x_1, \ldots, x_s \rightarrow 1}
{e^{-\hat\rho}\over e^{\rho_0} \R_0}
\sum_{\alpha \in M^\sharp} \;
\sum_{w \in W}
\epsilon (w) w \, t_\alpha
{e^{\hat\rho} \over \prod_j (1 + x_j e^{-\beta_j} )}.
\leqno(4.5)
$$
Assuming that $ |x_j|\/$ are sufficiently small we expand in a
geometric progression for each $ \alpha\/$:
$$
t_\alpha { e^{\hat\rho} \over \prod_j (1 + x_j e^{-\beta_j})} =
q^{{1\over2} h^\vee (\alpha|\alpha) + (\rho|\alpha)}
\sum_{\vec n \in \ZZ^s_+}
\prod_j
\left(
   -x_j q^{-(\alpha | \beta_j)}
\right)^{n_j} e^{\hat\rho + h^\vee \alpha - \sum_j n_j \beta_j}.
$$
Hence we may rewrite (4.5) as follows:
$$
\eqalign{
{| W / W^\sharp| \widehat\R \over e^{\rho_0} \R_0}
     & = \lim_{x_1, \ldots, x_s \rightarrow 1} \sum_{\alpha\in M^\sharp}
          q^{{1\over2} h^\vee (\alpha|\alpha) + (\rho|\alpha)} \cr
     & \qquad \times \sum_{\vec n \in \ZZ^s_+} \prod_j (-x_j q^{-
          (\alpha|\beta_j)})^{n_j} {\sum_{w \in W} \epsilon(w)
          e^{w (\hat\rho + h^\vee \alpha - \sum_j n_j \beta_j) -
          \hat\rho} \over e^{\rho_0} \R_0} .
}% End of eqalign
\leqno(4.6)
$$
Applying the specialization $ F_{t \rho_0}\/$ to both sides and
tending $ t\/$ to 0 we obtain:
$$
\displaylines{
\qquad\qquad
{| W / W^\sharp| \over 2^{N_1}} \prod_{n=1}^\infty
{(1 - q^n)^{\dim \gg_0} \over (1 + q^n)^{\dim \gg_1}} \hfill \cr
= \lim_{x_1, \ldots, x_s \rightarrow 1} \sum_{\alpha \in M^\sharp}
   q^{{1\over2} h^\vee (\alpha|\alpha) + (\rho|\alpha)}
   \sum_{\vec n \in \ZZ^s_+} \prod_j
   \left(
      -x_j q^{- (\alpha | \beta_j)}
   \right)^{n_j} \prod_{\gamma \in \Delta_{0+}} {(\rho + h^\vee
      \alpha - \sum_j n_j \beta_j | \gamma) \over (\rho_0 |
      \gamma)}.
}% End of displaylines
$$
The theorem now follows from (3.7).
\endproof

\beginproclaim{Theorem}{3}
Let $ \gg\/$ be a simple finite-dimensional Lie superalgebra of type
$ A (m, n)\/$, $ m \neq n\/$, or $ C(m)\/$.  Then in the notation of
Theorem 4.2 one has:
$$
\prod^\infty_{j=1} {(1-q^j)^{d_0} \over (1 + q^{j -
{1\over2}} )^{d_1}}
   = {q^{r + {1\over2} s} \over |W / W^\sharp |}
     \sum_{\alpha \in M^\sharp}
     q^{{1\over2} h^\vee (\alpha|\alpha) + (\rho + \beta|\alpha)}
\sum_{\vec n \in \ZZ^s_+ \atop |\vec n| \leq
     N_0} {A_\alpha ( \vec n) \over \prod_i
        \left(1 + q^{(\alpha|\beta_i)
             + {1\over2}}
        \right)^{n_i + 1}},
\leqno(4.7)
$$
where $ r\/$ is the number of roots from $ \Delta_{1+}\/$ which are
negative with respect to the standard set of positive roots.
\endproclaim

\beginproof
Let $ \lambda \in \hh^\ast\/$ be such that $ (\lambda | \alpha) =
0\/$ if $ \alpha \in \Delta_0\/$ and $ (\lambda | \alpha) =
{1\over2}\/$ if $ \alpha\/$ is odd and positive with respect to the
standard set of positive roots.  Now apply to both sides of (4.6)
the specialization $ F_{t \rho_0 - (\log q) \lambda}\/$ and tend $
t\/$ to 0.
\endproof

\beginexample{1}
Let $ \gg = sl (2, 1)\/$.  Choose $ \Delta_+\/$ such that both
simple roots $ \alpha_1\/$ and $ \alpha_2\/$ are odd.  Then $
(\alpha_1 | \alpha_1) = (\alpha_2 | \alpha_2 ) = 0\/$, $
(\alpha_1 | \alpha_2 ) = 1\/$, $ \rho = 0\/$, $ h^\vee = 1\/$, $
M^\sharp = \ZZ (\alpha_1 + \alpha_2)\/$.  We let $ x =
e^{-\alpha_1}\/$,
$ y = e^{-\alpha_2}\/$.  Then (2.4) becomes (if we take $ \beta =
\alpha_1\/$ or $ \alpha_2\/$ respectively):
$$
R := {1 - xy \over (1 + x) (1 + y)} =
{1 \over 1 + x} - {y \over 1 + y} =
{1 \over 1 + y} - {x \over 1 + x}.
$$
Formula (4.2) for $ \widehat\R\/$ becomes after expanding the
right-hand side (for which we assume $ |q| < |x|\/$, $ |y| < 1\/$):
$$
\matrix{
   \lefteqn{\prod_{n=1}^\infty
   {(1 - q^n)^2 (1 - xyq^{n-1}) (1 - x^{-1} y^{-1} q^n)
   \over
   (1 + xq^{n-1}) (1 + x^{-1} q^n) (1 + yq^{n-1}) (1 + y^{-1} q^n)}
   }
   & & \cr
   & = &
   \displaystyle
   \left(
     \sum^\infty_{m,n=0} - \sum^{-\infty}_{m,n=-1}
   \right)
     \left(
       (-1)^{m+n} x^m y^n q^{mn}
     \right) .
   \cr
}% End of matrix
\leqno(4.8)
$$
This truly remarkable identity may be viewed also as the denominator
identity for $ N=4\/$ superconformal algebra.  An equivalent form of
(4.8) is (here we assume $ |q| < |x| < 1)\/$:
$$
\prod^\infty_{k=1}
{(1 - q^n)^2 (1 - x y q^{n-1}) (1 - x^{-1} y^{-1} q^n)
\over
(1 + x q^{n-1}) (1 + x^{-1} q^n) (1 + y q^{n-1}) (1 + y^{-1} q^n)}
  =
\sum_{j \in \ZZ} { (-x)^j \over 1 + y q^j}.
$$
Replacing in this formula $ q\/$ by $ q^2\/$, $ x\/$ by $ z\/$ and
$ y\/$ by $ q z^{-1}\/$, we obtain
$$
\prod^\infty_{n = 1}
{(1 - q^n)^2
\over (1 + q^{n-1} z) (1 + q^n z^{-1})} =
\sum_{n \in \ZZ} (-1)^n
{q^{n(n+1)/2} \over 1 + q^n z} .
$$
This identity may be viewed as the denominator identity for $ N =
2\/$ superconformal algebra and also of the affine superalgebra $ gl
(1,1)^\wedge\/$.  It also has been rediscovered many times (see,
{\it e.g.,} [KP]).  It is worth mentioning here that Gauss identities
(5.1) and (5.2) are denominator identities for $ N = 1\/$
superconformal algebras.

Yet another useful form of (4.8) is obtained by dividing both sides
by $ \R\/$:
$$
\matrix{
\lefteqn{\displaystyle
  \prod^\infty_{n=1}
  {(1 - q^n)^2 (1 - x y q^n) (1 - x^{-1} y^{-1} q^n)
   \over (1 + x q^n ) (1 + x^{-1} q^n) (1 + y q^n) (1 + y^{-1}
   q^n)} \qquad} \cr
& = & \displaystyle
1 +
{(1 + x) (1 + y)
\over (1 - xy)}
\sum^\infty_{m,n=1} (-1)^{m+n} q^{mn} (x^m y^n - x^{-m} y^{-n}). \cr
}% End of matrix
\leqno(4.9)
$$
Letting $ x = y = z\/$, we obtain:
$$
\matrix{
\lefteqn{
   \displaystyle
   \prod^\infty_{n=1}
   {(1 - q^n)^2 (1 - z^2 q^n) (1 - z^{-2} q^n)
   \over (1 + z q^n )^2 (1 + z^{-1} q^n)^2} \qquad} \cr
  & = & \displaystyle
1 +
(1 + z) \sum^\infty_{m,n=1} (-1)^{m+n} q^{mn}
{z^{m+n} - z^{-m-n}
\over 1 - z} . \cr
}% End of matrix
\leqno(4.10)
$$
\endexample

\beginexample{2}
Let $ \gg = \osp (3,2)\/$.  Choose $ \Delta_+\/$ such that both
simple roots $ \alpha_1\/$ and $ \alpha_2\/$ are odd; one has: $
(\alpha_1 | \alpha_2 ) = 0\/$, $ (\alpha_2 | \alpha_2 ) =
{1\over2}\/$, $ (\alpha_1 | \alpha_2) = -{1\over2}\/$, $ \rho = -
{1\over2} \alpha_1\/$, $ h^\vee = {1\over2}\/$, $ M^\sharp = 2 \ZZ
\alpha_2\/$.  We let $ x = e^{-\alpha_1}\/$, $ y = e^{-\alpha_2}\/$.
Then (2.4) becomes:
$$
\R :=
{(1 - y) (1 - x y)
\over (1 + x) (1 + x y^2)} =
{1 \over 1 + x} - {y \over 1 + x y^2}.
$$
Formula (4.2) for $ \widehat\R\/$ becomes after expanding the
right-hand
side (as before, $ |q| < |x|\/$, $ |y| < 1\/$):
$$
\vbox{\halign{\hfil$\displaystyle #$\hfil \cr
\prod^\infty_{n=1}
{(1 - q^n)^2 (1 - x y q^{n-1}) (1 - x^{-1} y^{-1} q^n) (1 - y^2
q^{n-1}) (1 - y^{-2} q^n)
\over (1 + x q^{n-1}) (1 + x^{-1} q^n) (1 + y q^{n-1}) (1 + y^{-1}
q^n) (1 + x y^2 q^{n-1}) (1 + x^{-1} y^{-2} q^n)} \cr
=
\left(
\sum^\infty_{m,n=0} - \sum^{-\infty}_{m,n = -1}
\right)
( - x)^n
\left(
y^{-m} - y^{m + 2n + 1}
\right) q^{{1\over2} m (m + 2n + 1)}. \cr
}% End of halign
}% End of vbox
\leqno(4.11)
$$
\endexample

Here is a general construction of a specialization ({\it cf.} [K9]).
Let $ \sigma\/$ be an automorphism of order $ m\/$ of $ \gg\/$ such
that $ \sigma (h) = h\/$ for $ h \in \hh\/$, let $ \gg =
\bigoplus_{j \in \ZZ / m \ZZ} \gg_{\epsilon, j}\/$ ($ \epsilon =
0\/$ or 1) be the corresponding $ \ZZ / m \ZZ\/$-gradations and let
$ d_{\epsilon, j} = \dim \gg_{\epsilon, j}\/$.  For $ \alpha \in
\Delta\/$ we have $ \gg_\alpha \subset \gg_{\epsilon ,
\bar{s}_\alpha}\/$; let $ s_\alpha \in \ZZ\/$ be such that $ 0 \leq
s_\alpha < m\/$, $ \bar{s}_\alpha = s_\alpha \mod m\/$.  Define $
\lambda \in \hh\/$ by $ (\lambda | \alpha) = s_\alpha\/$.  Replace
$ q\/$ by $ q^m\/$ in (4.2), divide both sides of it by $
\prod_{\alpha \in \Delta_{0+} \atop (\lambda | \alpha) =
0} (1 - e^{-\alpha})\/$ and apply to the resulting identity the
specialization $ F^\sigma := F_{-(\log q) \lambda}\/$.  The
left-hand side becomes
$$
\prod_{j \in \NN} (1 - q^j)^{d_{0, j}} / (1 + q^j)^{d_{1, j}}.
$$
One can generalize the trick applied in the proof of Theorems 4.2
and 4.3 to evaluate the right-hand side (as in [K9]).  However,
the answer is simpler when $ \lambda\/$ is fixed under the
action of $ W\/$ (as in Theorems 4.2 and 4.3).  Another situation
when this trick may be applied is when $ s_\alpha = 0\/$ for $
\alpha \in \Delta^\sharp_0\/$ and $ s_\alpha \neq 0\/$ for $ \alpha
\in \Delta_0 \setminus \Delta^\sharp_0\/$, so that $ \lambda\/$ is
fixed under $ W^\sharp\/$.

\beginexample{3}
(a) Let $ \sigma\/$ be the automorphism of $ \gg\/$ such that $
\sigma |_{\gg_0} = 1\/$, $ \sigma |_{\gg_1} = -1\/$.  Then applying
$ F^\sigma\/$ to $ \widehat\R / \R^\sharp_0\/$ gives an expansion of
$ \prod_{j \in \NN} (1 - q^{2j})^{d_0} / (1 - q^{2j - 1})^{d_1}\/$,
which is $ \Delta (q)^{d_0}\/$ (see (5.2) below) provided that $ d_0
= d_1\/$.

(b) Let $ \gg = A (n, n-1)\/$ and choose $ \Delta_+\/$ such that $
\Pi\/$ consists of odd roots $ \{ \alpha_1, \ldots,
\alpha_{2n}\}\/$ (see Example 1.1).  Let $ \sigma\/$ be such that $
s_{\alpha_{2j}} = 1\/$ and $ s_{\alpha_{2j - 1}} = 0\/$ ($ j = 1,
\ldots, n\/$).  Then $ F^\sigma\/$ gives an expansion of $\squaretwo
(q)^{2n}\/$ over an $ n\/$-dimensional lattice.  Namely, using the
same method as that in the proof of Theorem~4.2, we obtain:
$$
\squaretwo (-q)^{2n} = 2^n
\sum_{\alpha \in M^\sharp} \;
\sum_{s \in \SS}
(-1)^{|s|}
q^{{n + 1 \over 2} (\alpha | \alpha) + (\alpha | \rho_0) - (\alpha
   | \tilde s)}
\prod^n_{j=1}
\left(
1 + q^{(n+1)(\alpha | \beta_j) - (\tilde s | \beta_j)}
\right)^{-1},
$$
where $ \SS\/$ is the set of all subsets in $ \Delta_{0+}\/$, and $
|s|\/$ (resp.\ $ \tilde s\/$) stands for the number (resp.\ sum) of
elements of $ s \in \SS\/$.  Note that $ M^\sharp\/$ is an
$ n\/$-dimensional lattice (of type $ A_n\/$), so that for $ n =
1\/$ we recover Jacobi's formula (0.1), but for $ n =\/$ 2, 3 and 4
our formula is more complicated than Jacobi's (0.2--4).  The way to
recover Jacobi's formulas will be explained below.

(c) Let $ \gg = A (2n + 1, 2n) = sl (2n + 2, 2n + 1)\/$ and let $
\sigma\/$ be the conjugation by the matrix $ \diag (I_{n+1}, -
I_{n+1}\/$; $ I_{n+1}, -I_n)\/$.  Then $ \sigma\/$ is an order 2
automorphism of $ \gg\/$ such that $ d_{0, 0} = d_{0, 1} = d_{1, 0}
= d_{1, 1} = 2 (n+1) (2n +1)\/$.  Hence $ F^\sigma\/$ gives an
expansion of $ \squaretwo (q)^{2 (n+1)(2n+1)}\/$ over an
$ n+1\/$-dimensional lattice $ (n = 0, 1, \ldots)\/$.
\endexample

\beginexample{4}
Let $\frak g = A (m, n)\/$ with $m > n\/$.  We choose the set of
simple roots as in Example~1.1.  Then formula (4.2) can be
written, using (2.4) and (3.4), as follows:
$$
e^{\hat\rho} \hat \R = {1 \over (n+1)!} \sum_{\alpha \in M^\sharp}
\sum_{w \in W} \epsilon (w) w t_{-\alpha}
\left(
        e^{\hatrho} \prod^{n+1}_{j=1} \left(
        1 + e^{-\alpha_{2j - 1}} \right)^{-1}
\right),
\leqno(4.12)
$$
where $M^\sharp\/$ is the lattice spanned over $\ZZ\/$ by elements
$\alpha_{2j -1} + \alpha_{2j}\/$ ($j = 1, \ldots, n+1\/$) and
$\alpha_i\/$ ($2n + 3 \leq i \leq m + n + 1\/$).  Introduce the
element $\xi \in \frak h\/$ by relations $( \xi | \alpha_j ) =
(-1)^j / 2\/$ for $1 \leq j \leq 2(n+1)\/$, $(\xi | \alpha_j) =
0\/$ for $2n + 3 \leq j \leq m + n + 1\/$.  Then $(\xi | \alpha )
= 0\/$ for all $\alpha \in \Delta_0\/$, hence $t_\xi\/$ commutes
with $W\/$.  Applying $t_\xi\/$ to both sides of (4.12) we obtain
(assuming $| q | < 1\/$):
$$
\eqalign{ e^{\hatrho + \beta} \hat\R_T & = {1 \over (n+1)!}
\left(\prod^{n+1}_{j=1} \left(\sum^\infty_{k_j, r_j = 0} -
\sum^{-\infty}_{k_j,r_j = -1} \right) \right)
(-1)^{\sum^{n+1}_{j=1} r_j} \cr
& \qquad \times \sum_{k_{n+2}, \ldots, k_m \in \ZZ} \sum_{w \in W}
\epsilon(w) w e^{\hatrho - h^\vee \alpha - \sum^{n+1}_{j=1} r_j
\alpha_{2j - 1}} \cr
& \qquad \times q^{{1\over2} h^\vee (\alpha|\alpha) +
\sum^{n+1}_{j=1} r_j \left(k_j + {1 \over 2} \right) -
\sum^m_{j=n+2} k_j - {1\over4} (n+1) (2m-n)}. \cr
}% end of eqalign
\leqno(4.13)
$$
Here
$$
\hat\R_T = \R_0 \prod^\infty_{n=1}
{\prod_{\alpha \in \Delta_0} \left(1 - q^n e^\alpha \right)
  \over
\prod_{\alpha \in \Delta_1}\left(1 + q^{n - {1\over2}} e^\alpha
\right)},
$$
$\beta\/$ is the sum of the $\gamma \in \Delta_{1+}\/$ such that
$\left(\gamma | \xi \right) = {1\over2}\/$,
$$
\alpha = \sum^{n+1}_{j=1} k_j \left(\alpha_{2j-1} + \cdots +
\alpha_{2n+2} \right) + \sum^m_{j=n+2} k_j \alpha_{n+1+j}
$$
(so that $\left(\hatrho | \alpha \right) = \sum^m_{j=n+2}
k_j\/$).  Dividing both sides of (4.13) by $\R_0\/$ and letting
$e^{\alpha_i} \mapsto 1\/$ ($i = 1, \ldots, m + n + 1\/$), we
obtain (using the Weyl dimension formula) a power series
expansion for the product $\prod^\infty_{j=1} \left(1 - q^j
\right)^{d_0} / \left(1 - q^{j - {1 \over 2}} \right)^{d_1}\/$.
Furthermore, applying $t_{-\xi}\/$ to both sides of (4.13), we
obtain a better expression for the affine denominator: %
$$
\eqalign{
e^{\hatrho} \hat\R & = {1 \over (n+1)!}
\sum_{J \subset \left\{1, \ldots, n+1 \right\}}
\sum_{\left(\vec k, \vec r \right) \in S_J}
\sum_{k_{n+2}, \ldots, k_m \in \ZZ} \epsilon_J \bigl(\vec k, \vec
r \bigr) (-1)^{\sum_{j \in J} r_j} \cr
& \quad \times
\sum_{w \in W} \epsilon (w) w
{e^{\hatrho - h^\vee \alpha - \sum_{j \in J} r_j \alpha_{2j - 1}}
\over
\prod_{j \in J^c} \left(1 + e^{-\alpha_{2j - 1}} \right)}
q^{{1\over2} h^\vee (\alpha|\alpha) + \sum_{j\in J} r_j k_j -
\sum^m_{j=n+2} k_j} . \cr
}% end of eqalign
\leqno(4.14)
$$
Here
$$
S_j = \left\{\bigl( \left. \vec k, \vec r \bigr)_{j\in J} \in
\ZZ^{2\left|J \right|} \right| k_j > 0 \hbox{ and } r_j \geq 0,
\hbox{ or } k_j < 0,\; r_j < 0 \hbox{ for each } j \in J
\right\},
$$
$$
\epsilon_J \bigl(\vec k, \vec r \bigr) = (-1)^{\sharp\left\{j \in J
| k_j < 0 \right\}} \hbox{ and } J^c = \left\{1, 2, \ldots, n+1
\right\} \setminus J.
$$
Note that $\left(\rho -h^\vee \alpha - \sum_{j \in J} r_j
\alpha_{2j-1} | \alpha_{2i - 1} \right) = 0\/$ if $i \in J^c\/$.
Hence, by dividing both sides of (4.14) by $\R_0\/$ and letting
$e^{\alpha_i} \mapsto 1\/$ ($i = 1, \ldots, m+n+1\/$), we may use
Theorem~3.2 to obtain a power series expansion for
$\prod^\infty_{j=1} \left(1 - q^j \right)^{d_0} / \left(1 + q^j
\right)^{d_1}\/$.
\endexample

\beginexample{5}
Let $\frak g = B (m, m)\/$.  We choose a set of simple roots
$\alpha_1, \ldots, \alpha_{2m} \in \Delta_1\/$ with the following
scalar products:  $\left(\alpha_i | \alpha_i \right) = 0\/$ for
$i = 1, \ldots, 2m - 1\/$, $\left(\alpha_{2m} | \alpha_{2m}
\right) = - 1/2\/$, $\left(\alpha_i | \alpha_{i+1} \right) =
(-1)^{i + 1} / 2\/$ for $i = 1, \ldots, 2 m - 1\/$.  We have:
$h^\vee = 1/2\/$ and $M^{\sharp}\/$ is spanned over $\ZZ\/$ by
vectors $2 \left(\alpha_{2j - 1}+ \alpha_{2j} \right)\/$, $j = 1,
\ldots, m\/$.  Then formula (4.2) can be written using (2.4) and
(3.4), as follows:
$$
e^{\hatrho} \hat\R = {1 \over 2^m m!}
\sum_{\alpha \in M^\sharp}
\sum_{w \in W}
\epsilon (w) w t_{-\alpha} \left( e^{\hatrho} \prod^m_{j=1}
\left(1 + e^{-\alpha_{2j-1}} \right)^{-1} \right).
\leqno(4.15)
$$
Introduce the element $\xi \in \hh\/$ by relations $\left(\xi |
\alpha_j \right) = (-1)^j / 2\/$ for $j = 1, \ldots, m-1\/$,
$\left(\xi | \alpha_{2m} \right) = 0\/$.  Then $w (\xi) - \xi \in
M^{\sharp}\/$ for all $w \in W\/$.  Applying $t_\xi\/$ to both sides
of (4.15) we obtain a more convenient for specializations
formula:
$$
\eqalign{
\rlap{$\displaystyle
  e^{\hatrho + \beta} \hat\R_T = {1 \over 2^m m!}
  \left(\prod^m_{j=1} \left( \sum^\infty_{k_j, r_j = 0} -
  \sum^{-\infty}_{k_j, r_j = -1} \right) \right)
  (-1)^{{1 \over 2} m (m+1) + \sum^m_{j=1} r_j}$} \qquad\qquad & \cr
& \times \sum_{w \in W} \epsilon (w) w e^{\hatrho + {1 \over 2}
(\xi - \alpha) - \sum^m_{j=1} r_j \alpha_{2j-1}} \cr
& \times q^{{1 \over 4} (\alpha | \alpha) + \sum^m_{j=1}
\left(k_j (r_j+1) + {1 \over 2} r_j \right) - {1 \over 4} m (m-1)},
}% end of eqalign
\leqno(4.16)
$$
where $\beta = {1 \over 2} \xi +
\sum_{\gamma \in \Delta_{1+} : (\xi | \gamma) = {1\over2}} \gamma
- \sum_{\gamma \in \Delta_{0+} : (\xi | \gamma) = -1} \gamma \/$
and $\alpha = 2 \sum^m_{j=1} k_j \sum^{2m}_{i = 2j - 1}
\alpha_i\/$.
\endexample

\beginsection{5}{Applications to sums of squares and sums of
triangular numbers}

Consider the following power series in $ q\/$ (that converge for $
|q| < 1\/$):
$$
\squaretwo (q) = \sum_{n \in \ZZ} q^{n^2}, \quad
\Delta (q) = \sum_{n \in \ZZ_+} q^{n (n + 1) / 2} .
$$
One has the following product expansions due to Gauss (which follow
immediately from the Jacobi triple product identity):
$$
\leqalignno{
\squaretwo ( -q ) & = \prod_{k \geq 1} {1 - q^k \over 1 + q^k},
     & (5.1) \cr
\Delta (q) & = \prod_{k \geq 1} {1 - q^{2k} \over 1 - q^{2k-1}}.
     & (5.2) \cr
}% End of leqalignno
$$
Let $ \Delta_j = j ( j + 1) / 2\/$ and let $ \Delta = \{
\Delta_j | j \in \ZZ_+\}\/$ be the set of triangular numbers.
Let $ \squaretwo_{N, k} = | \{ \vec n \in \ZZ^N | \sum_{j=1}^N
n^2_j = k \} |\/$ (resp.\ $ \Delta_{N, k} = | \{ \vec n \in
\Delta^N | \sum^N_{j=1} n_j = k \} |\/$) be the number of
representations of $ k\/$ as a sum of $ N\/$ squares (resp.\
triangular numbers).  Let $ \squaretwo_N (q) = \sum_{k \in \ZZ_+}
\squaretwo_{N, k} q^k\/$ and $ \Delta_N (q) = \sum_{k \in \ZZ_+}
\Delta_{N, k} q^k\/$ be the generating series for these numbers.
Then, clearly %
$$
\squaretwo_N (q) = \squaretwo (q)^N, \quad
\Delta_N (q) = \Delta (q)^N.
$$
Thus, in view of (5.1) and (5.2) we may use the results of Section~4
to obtain expansion for $ \squaretwo_N (q)\/$ and $ \Delta_N (q)\/$.
In fact,  formulas (4.3) and (4.7) give such expansions provided
that $ d_0 = d_1\/$, which happens for $ \gg\/$ of type $ A ( n,
n - 1)\/$, $ B (n, n)\/$ or $ D (n + 1, n+ 1)\/$.  These gives us
formulas for $ \squaretwo_N (q)\/$ and $ \Delta_N (q)\/$, where
respectively
$$
N = 2n (n + 1),\; 2n (2n +1) \hbox{ or } 4 (n + 1)^2 \qquad (n \in
\NN).
$$

\beginexample{1}
$ \gg = A(1,0)\/$.  Letting $ z \rightarrow 1\/$ in (4.10) we get:
$$
\eqalign{
\squaretwo ( -q )^4
     &= 1 - 4 \sum^\infty_{m,n=1} (-1)^{m+n} (m + n) q^{mn} \cr
     &= 1 + 8 \sum^\infty_{m,n=1} (-1)^{m+n-1} m q^{mn} \cr
     &= 1 + 8 \sum^\infty_{m=1} {m (-q)^m \over 1 + q^m}.
}% End of eqalign
$$
Since $ {2 m q^{2m} \over 1 + q^{2m}} =
{2 m q^{2m} \over 1 - q^{2m}} -
{4m q^{4m} \over 1 - q^{4m}}\/$, we obtain:
$$
\squaretwo (q)^4 = 1 + 8 \sum_{k \geq 1 \atop 4 \nmid k}
{k q^k \over 1 - q^k},
$$
which is Jacobi's formula (0.2).  In other words, for $ n \geq 1\/$:
$$
\squaretwo_{4,n} = 8 \sum_{k | n \atop 4 \nmid k} k.
$$

Letting $ z = i\/$ in (4.9), we get Jacobi's formula (0.1), {\it
i.e.,\/} that $ \squaretwo_{2, n}\/$ equals 4 times the difference
between the number of divisors of $ n\/$ congruent to 1 mod 4 and
the number of divisors of $ n\/$ congruent to $ -1\/$ mod 4.  This
result goes back to Gauss.

Letting in (4.9) $ x = -y = q^{1/2}\/$, we obtain:
$$
\Delta (q)^2 = \sum^\infty_{j, k = 0} (-1)^k
q^{\left( ( 2j + 1) (2k + 1) - 1\right) / 4}.
$$
This means that $ \Delta_{2,n}\/$ equals the difference between
the number of divisors of $ 4n + 1\/$ congruent to 1 mod 4 and the
number of divisors of $ 4n + 1\/$ congudent to $ -1\/$ mod 4.
Replacing in (4.9) $ q\/$ by $ q^2\/$, letting $ x = - qz\/$, $ y
= -q\/$ and tending $ z\/$ to 0, we get
$$
\Delta (q)^4 = \sum^\infty_{j,k=0} (2k + 1)
q^{\left( ( 2j + 1 ) ( 2k + 1 ) -1 \right) / 2}.
$$
This means that $ \Delta_{4,n}\/$ equals the sum of all divisors
of $ 2n + 1\/$.  This result is due to Legendre (1828).
\endexample

\beginexample{2}
$ \gg = B (1,1)\/$.  Dividing both sides of (4.11) by $ (1 - xy) (1
- y)\/$ and letting $ x = y = z \rightarrow 1\/$, we obtain Jacobi's
formula (0.3).  In other words,
$$
\squaretwo_{6,n} = 4
\sum_{j,k \geq 1 \atop j {\rm odd}, j\kern-1pt k = n}
(-1)^{{1\over2} (j+1)} (j^2 - 4k^2).
$$
Furthermore, replacing in (4.11) $ q\/$ by $ q^2\/$ and letting $ x
= y = -qz\/$, we obtain after dividing both sides by $ (z - 1)^2\/$
and letting $ z \rightarrow 1\/$:
$$
\Delta (q)^6 = - {1 \over 8}
\sum_{{j,k \, {\rm odd} \atop j > k \geq 1} \atop {1 \over 2}
(j - k) \rm odd}
(-1)^{{1 \over 4} (j-1) (k+1)} (j^2 - k^2) q^{{1 \over 4} (jk - 1)}.
$$
In other words, as can be easily seen:
$$
\Delta_{6,n} = {1 \over 8}
\sum_{j > k \geq 1 \atop j \kern-1pt k = 4n + 3}
(-1)^{{1 \over 2} (j + 1)} \left(j^2 - k^2\right).
$$

\beginexample{3}
$\frak g = A (m, m-1)\/$.  Dividing both sides of (4.13) by
$\R_0\/$, letting $e^{\alpha_i} \rightarrow 1\/$ ($i = 1, \ldots,
2m\/$) and replacing $q^{1\over2}\/$ by $-q\/$, we obtain
$$
\eqalign{
  \Delta (q)^{2m (m+1)} & =
  \left(
    \sum_{\vec k, \vec r \in \ZZ^m_+} -
    \sum_{\vec k, \vec r \in (-\NN)^m}
  \right)
  \prod^m_{j=1}
  {r'_j + \bigl| \vec k \bigr| \over (j!)^2}
  \left(
    \prod_{1 \leq i < j \leq m} (r_i - r_j) (r'_i - r'_j)
  \right)\cr
  & \qquad \times
  q^{2 \sum_{1 \leq i \leq j \leq m} k_i k_j + \sum^m_{j=1} r_j
    (2 k_j + 1) - m (m+1) / 2},
}% end of eqalign
$$
where
$$
  r'_j = r_j + k_j .
  \leqno(5.3)
$$
Similarly, we obtain from (4.14) using the remarks and notation
of Example 4.4:
$$
\eqalign{
  \squaretwo (-q)^{2m (m+1)} & =
  {(-1)^{{1 \over 2}m (m+1)} 2^m \over
    \prod^m_{j=1} (j!)^2}
  \sum_{J \subset \left\{1, \ldots, m \right\}}
  \sum_{\left(r_j \right)_{j\in J^c}}
  \sum_{\bigl(\vec k, \vec r \bigr) \in S_J}
  \epsilon_J \bigl(\vec k, \vec r \bigr) (-1)^{\sum^m_{j=1} r_j}
  \cr
  & \quad
  \times 2^{-\left|J^c \right|} C
  \left(
    \sum_{i \in J^c}
    r_i +
    \left| J^c
    \right|
    - 1,\, m^2 +
    \left| J^c
    \right| -1
  \right)
  \prod^m_{j=1}
  \left( r'_j + k_1 + \cdots + k_m
  \right) \cr
  & \quad
  \times \prod_{1 \leq i < j \leq m} \left(r_i - r_j \right)
  \left(r'_i - r'_j
  \right)
  q^{\sum_{1 \leq i \leq j \leq m} k_i k_j + \sum_{j \in J} k_j
    r_j}
  }% end of eqalign
$$
\endexample

\beginexample{4}
$\gg = B (m, m)\/$.  Dividing both sides of (4.16) by $\R_0\/$,
letting $e^{\alpha_i} \rightarrow 1\/$ ($i = 1, \ldots, 2m\/$)
and replacing $q^{1 \over 2}\/$ by $-q\/$ we obtain (using the
Weyl dimension formula):
$$
\eqalign{
  \Delta (q)^{2 m (2m + 1)} & = {1 \over 2^m m!}
  \left(
    \sum_{\vec k, \vec r \in \ZZ^m_+} -
    \sum_{\vec k, \vec r \in (-\NN)^m}
  \right)
  \prod^m_{j=1}
  {(-1)^{k_j} \over \left((2j - 1)! \right)^2}
  (1 + 2 r_j) (1 + r'_j)\cr
  & \qquad \times
  \prod_{1 \leq i < j \leq m}
  \left(
    r_i - r_j
  \right)
  \left(
    r'_i - r'_j
  \right)
  \left(
    1 + r_i + r_j
  \right)
  \left(
    2 + r'_i + r'_j
  \right) \cr
  & \qquad \times
  q^{\sum^m_{j=1}
    \left(
      k^2_j + 2 r_j k_j + 2 k_j + r_j
    \right) - {1 \over 2} m (m-1)} .
  }% end of eqalign
$$
\endexample

\beginconjecture{1}
$$
\eqalign{
  \Delta (q)^{2 m (2m + 1)} & =
  {1 \over m!
    \left(
      \prod^m_{j=1} (2j - 1)!
    \right)^2
    }% end of over
  \sum_{\vec k, \vec r \in \ZZ^m_+} (-1)^{\left|\vec k \right|} \cr
  & \quad \times
  \prod_{1 \leq i < j \leq m}
  \left(
    r_i - r_j
  \right)
  \left(
    r'_i - r'_j
  \right)
  \left(
    2 + r'_i + r'_j
  \right)
  \left(
    1 + r_i + r_j
  \right) \cr
  & \quad \times
  \prod^m_{j = 1}
  \left(
    1 + 2 r_j
  \right)
  \left(
    1 + r'_j
  \right)
  q^{\sum^m_{j = 1}
    \left(
      \left(
        k^2_j + 2 k_j
      \right) +
      \left(
        r_j
        \left(
          2 k_j + 1
        \right)
      \right)
    \right)
    - {1 \over 2} m (m - 1) } . % end of superscript
  }% End of eqalign
$$
\endconjecture

\beginsection{6}{Some ``degenerate'' theta functions and modular
forms}

In this section we use the denominator identity (4.8) for $ sl
(2,1)^\wedge\/$ to derive transformation properties of certain
``degenerate'' theta functions and modular forms, and related to
them characters of $ N = 2\/$ superconformal algebras.

Let $ q = e^{2 \pi i \tau}\/$ where $ I m \tau > 0\/$, and let $
\Theta (\tau, z)\/$ be the standard Jacobi theta function:
$$
\eqalign{
  \Theta (\tau, z) & =
  e^{\pi i \tau / 4}
  e^{- \pi i z}
  \prod_{n \geq 1}
  \left(
    1 - q^n
  \right)
  \left(
    1 - q^n e^{-2 \pi i z}
  \right)
  \left(
    1 - q^{n-1} e^{2 \pi i z}
  \right) \cr
  & =
  \sum_{j \in \ZZ}
  (-1)^j
  q^{{1 \over 2} (j - {1 \over 2})^2}
  e^{2 \pi i
    \left( j - {1 \over 2}
    \right) z}. \cr
  }% End of eqalign
$$
Recall that this theta function has the following transformation
property (see, {\it e.g.\/} [K9]):
$$
\Theta
\left(
     - {1 \over \tau}, {z \over \tau}
\right) = i (-i \tau)^{1/2} e^{\pi i z^2 / \tau} \Theta (\tau, z).
$$
Introduce the following function in $ \tau\/$, $ z_1\/$, and $
z_2\/$:
$$
F (\tau, z_1, z_2 ) =
{\eta (\tau)^3 \Theta (\tau, z_1 + z_2)
\over \Theta (\tau, z_1) \Theta (\tau, z_2)},
$$
where $ \eta (\tau)\/$ is the Dedekind $ \eta\/$-function.  Note
that $ F (\tau, z_1, z_2)\/$ is the left-hand side of identity (4.8)
where $ x = - e^{2 \pi i z_1}\/$, $ y = - e^{2\pi i z_2}\/$.

Fix a positive integer $ m\/$.  Using the transformation formula for
$ \Theta\/$ and identity (4.8), it is not hard to derive the
following transformation formula:
$$
F
\left(
     - {m \over \tau} , {z_1 \over \tau}, {z_2 \over \tau}
\right) = {\tau \over m} e^{2 \pi i z_1 z_2 \over m \tau}
\sum_{a, b \in \ZZ / m \ZZ}
e^{2 \pi i (a z_1 + b z_2 ) \over m} q^{a b \over m}
F ( m \tau, z_1 + a \tau, z_2 + b \tau ).
\leqno(6.1)
$$

Given $ j, k \in {1 \over 2} \ZZ\/$, introduce the following
functions:
$$
F^{(m)\pm}_{j,k} (\tau, z) =
q^{j k \over m} e^{2 \pi i (j - k) z \over m}
F
\left(
   m \tau, -z + j \tau - {\textstyle{ 1 \over 4} \pm { 1 \over 4}, z
   + k \tau + { 1 \over 4} \mp {1 \over 4}}
\right).
$$
We call them {\it degenerate\/} theta functions since, due to (4.8),
they have the following power series expansion associated to a
bilinear form representing zero:
$$
F^{(m)\pm}_{j,k} (\tau, z) =
\left(
     \sum^\infty_{a, b = 0} - \sum^{-\infty}_{a,b=-1}
\right) (\pm 1)^{b-a} q^{m \left(a + {k \over m}\right) \left(b + {j
\over m}\right)}
e^{2 \pi i \left( \left(b + {j \over m}\right) - (a + {k \over
   m }) \right) z}.
\leqno(6.2)
$$

Transformation formula (6.1) implies the transformation formulas for
indefinite theta functions.  In order to state them introduce the
following sets:
$$
\eqalign{
A^+_m & = \{ (j, k) \in \ZZ^2 | 0 < j, j + k < m, \quad 0 \leq k <
m \}, \cr
A^-_m & =
\left\{
   (j, k) \in
   \left(
      {\textstyle { 1 \over 2}} + \ZZ
   \right)^2 | 0 < j, k, j + k < m
\right\}. \cr
}%End of eqalign
$$
Introduce the following numbers:
$$
S^{(m)}_{(j,k) (a,b)} =
{2 \over m} e^{\pi i (j - k) (a - b) \over m}
\sin {\pi (j + k) (a + b) \over m}.
$$
Provided that $ j, k \in \ZZ\/$, one has:
$$
\leqalignno{
F^{(m)\pm}_{j, k} \left(- {1 \over \tau}, {z \over \tau} \right)
   & = - i \tau e^{- {2 \pi i z^2 \over m \tau}}
   \sum_{(a, b) \in A^\pm_m} S^{(m)}_{(j, k) (a, b)}
   F^{(m)+}_{a, b} (\tau, z). & (6.3) \cr
\noalign{\hbox{Provided that $ j, k \in {1 \over 2} + \ZZ\/$, one
has:}}
F^{(m)\pm}_{j, k} \left(- {1 \over \tau}, {z \over \tau} \right)
   & = - i \tau e^{- {2 \pi i z^2 \over m \tau}}
   \sum_{(a, b) \in A^\pm_m} S^{(m)}_{(j, k) (a, b)}
   F^{(m)-}_{a, b} (\tau, z). & (6.4) \cr
}% End of leqalignno
$$
It follows that the linear span of functions $ \{ F^{(m)+}_{j,k}\/$
and $ F^{(m)-}_{j,k} | (j,k) \in A^+_m \cup A^-_m \}\/$ is invariant
with respect to the action of $ SL_2 (\ZZ)\/$ given by
$$
F (\tau, z) |_{\left( {a \atop c} {b \atop d} \right)} =
( c \tau + d )^{-1} e^{2 \pi i c z^2 \over m (c \tau + d)} F
\left(
   {a \tau + b \over c \tau + d} \, ,
   {z \over c \tau + d}
\right).
$$

We want now to specialize to $ z = 0\/$.  Let $ \varphi^{(m)+}_{j,k}
(\tau) = F^{(m)+}_{j,k} (\tau, 0)\/$ for all $ (j, k) \in A^-_m\/$
and $ \varphi^{(m)-}_{j,k} (\tau) = F^{(m)-}_{j,k} (\tau, 0)\/$ for
all $ (j,k) \in A^-_m \cup A^+_m\/$ except for $ (j, 0) \in A^+_m\/$
(when the series diverges).  These functions have the following
power series expansions:
$$
\varphi^{(m)\pm}_{j,k} (\tau) =
\left(
   \sum^\infty_{a,b=0} - \sum^{-\infty}_{a,b=-1}
\right)
(\pm 1)^{a+b} q^{m \left( a + {j \over m}\right) \left( b + {k \over
m}\right)} .
$$
In order to get $ SL_2 (\ZZ)\/$-invariance, we must add functions
$ \varphi^{(m)-}_{j,0} (\tau)\/$ ($j \in \ZZ\/$, $ 0 < j < m\/$)
given by
$$
\varphi^{(m)-}_{j,0}  = {\textstyle {1 \over 2}} +
\left(
   \sum^\infty_{a,b = 0 \atop b > 0} - \sum^{-\infty}_{a,b = -1}
\right)
   (-1)^{a + b} q^{m \left( a + {j \over m} \right) b}.
$$
Then we have:
$$
\eqalign{
\varphi^{(m)+}_{j,k} \left( - {1 \over \tau} \right)
   & = - i \tau \sum_{(a, b) \in A^+_m} S_{(j, k) (a, b)}
      \varphi^{(m)-}_{a, b} (\tau) \hbox{ if } (j, k) \in A^-_m,
      \cr
\varphi^{(m)-}_{j,k} \left( - {1 \over \tau} \right)
   & = - i \tau \sum_{(a, b) \in A^-_m} S_{(j, k) (a, b)}
      \varphi^{(m)-({\rm resp}+)}_{a, b} (\tau) \hbox{ if } (j, k)
      \in A^{-({\rm resp}+)}_m. \cr
}% End of eqalign
$$
Note also that
$$
\eqalign{
  \varphi^{(m)+}_{j,k} (\tau + 1)
  & = e^{2 \pi i j k \over m} \varphi^{(m)-}_{j,k} (\tau)
  \hbox{ if } (j, k) \in A^-_m, \cr
  \varphi^{(m)-}_{j,k} (\tau + 1)
  & = e^{2 \pi i j k \over m} \varphi^{(m)+ ({\rm resp} -)}_{j,k}
  (\tau)
  \hbox{ if } (j, k) \in A^{- ({\rm resp} +)}_m. \cr
  }% End of eqalign
$$
Thus we have constructed a finite set of modular forms of weight 1
whose linear span is $ SL_2 (\ZZ)\/$-invariant.

We derive now, using (6.3) and (6.4) the transformation properties
of characters and supercharacters of the $ N = 2\/$ superalgebras
({\it cf.\/} [RY]).  Introduce the following normalized $ N =
2\/$ demoninators:
$$
\eqalign{
\Psi^{(0)\pm} (\tau, z)
   & = e^{\pi i z} \prod^\infty_{n = 1}
      { (1 - q^n)^2 \over (1 \pm q^n e^{-2 \pi i z}) (1 \pm q^{n-1}
      e^{2 \pi i z})}, \cr
\Psi^{(1/2)\pm} (\tau, z)
   & = q^{1/8} \prod^\infty_{n=1} { (1 - q^n )^2 \over (1 \pm q^{n
   - 1/2} e^{- 2 \pi i z}) (1 \pm q^{n - 1/2} e^{2 \pi i z})}.
}% End of eqalign
$$
{}From the transformation formula for $ \Theta\/$ and $ \eta\/$ one
gets:
$$
\eqalign{
\Psi^{(0)-} \left( - {1 \over \tau}, {z \over \tau} \right)
   & = \tau e^{- {\pi i z^2 \over \tau}} \Psi^{(0)-} (\tau, z), \cr
\Psi^{(0)+} \left( - {1 \over \tau}, {z \over \tau} \right)
   & = -i \tau e^{- {\pi i z^2 \over \tau}} \Psi^{(1/2)-} (\tau,
      z), \cr
\Psi^{(1/2)-} \left( - {1 \over \tau}, {z \over \tau} \right)
   & = -i \tau e^{- {\pi i z^2 \over \tau}} \Psi^{(0)+} (\tau,
      z), \cr
\Psi^{(1/2)+} \left( - {1 \over \tau}, {z \over \tau} \right)
   & = -i \tau e^{- {\pi i z^2 \over \tau}} \Psi^{(1/2)+} (\tau,
      z), \cr
}% End of eqalign
$$

One knows that the normalized characters and supercharacters of the
unitary discrete series of the two $ N = 2\/$ superalgebras
corresponding to $ \epsilon = 0\/$ or $ 1 \over 2\/$ of central
charge $ 3 c_m\/$, where $ c_m = 1 - {2 \over m}\/$, $ m = 2, 3,
\ldots,\/$ are given by the following formulas
([D],[M],[RY]):
$$
\eqalign{
  \ch^{(m,\epsilon)}_{j,k} (\tau, z)
  & =
  {F^{(m)-}_{j,k} (\tau, z)
    \over \Psi^{(\epsilon)+} (\tau, z)}, \cr
  \sch^{(m,\epsilon)}_{j,k} (\tau, z)
  & =
  {F^{(m)+}_{j,k} (\tau, z)
    \over \Psi^{(\epsilon)-} (\tau, z)}, \cr
  }% End of eqalign
$$
if $ j, k \in \epsilon + \ZZ\/$; the range of $ (j, k)\/$ is $
A^+_m\/$ or $ A^-_m\/$ depending on whether $ \epsilon = 0\/$ or $
1 \over 2\/$.  From the above transformation formulas we obtain the
following theorem.

\beginproclaim{Theorem}{1}
({\it cf.\/} [RY]) (a) If $ (j, k) \in A^-_m$, then
$$
\eqalign{
\ch^{(m, 1/2)}_{j,k} \left( - {1 \over \tau}, {z \over \tau} \right)
   & = e^{{\pi i z^2 \over \tau} c_m} \sum_{a, b \in A^-_m}
      S^{(m)}_{(j, k) (a, b)} \ch^{(m, 1/2)}_{a,b} (\tau, z), \cr
\sch^{(m, 1/2)}_{j,k} \left( - {1 \over \tau}, {z \over \tau}
   \right)
   & = e^{{\pi i z^2 \over \tau} c_m} \sum_{a, b \in A^+_m}
      S^{(m)}_{(j, k) (a, b)} \ch^{(m, 0)}_{a,b} (\tau, z), \cr
\ch^{(m, 1/2)}_{j,k} \left( \tau + 1, z \right)
   & = e^{{2 \pi i j k \over m} - {\pi i \over 4}} \sch^{(m,
      1/2)}_{j, k} (\tau, z),
      \cr
\sch^{(m, 1/2)}_{j,k} \left( \tau + 1, z \right)
   & = e^{{2 \pi i j k \over m} - {\pi i \over 4}} \ch^{(m,
      1/2)}_{j,k} (\tau, z)
      . \cr
}% End of eqalign
$$

(b) If $ (j, k) \in A^+_m$, then
$$
\eqalign{
\ch^{(m, 0)}_{j,k} \left( - {1 \over \tau}, {z \over \tau} \right)
   & = e^{{\pi i z^2 \over \tau} c_m} \sum_{a, b \in A^-_m}
      S^{(m)}_{(j, k) (a, b)} \sch^{(m, 1/2)}_{a,b} (\tau, z), \cr
\sch^{(m, 0)}_{j,k} \left( - {1 \over \tau}, {z \over \tau} \right)
   & = - i e^{{\pi i z^2 \over \tau} c_m} \sum_{a, b \in A^+_m}
      S^{(m)}_{(j, k) (a, b)} \sch^{(m, 0)}_{a,b} (\tau, z), \cr
\ch^{(m, 0)}_{j,k} \left( \tau + 1, z \right)
   & = e^{2 \pi i j k \over m} \ch^{(m, 0)}_{j, k} (\tau, z),
      \cr
\sch^{(m, 0)}_{j,k} \left( \tau + 1, z \right)
   & = e^{{2 \pi i j k \over m}} \sch^{(m, 0)}_{j,k} (\tau, z)
      . \cr
}% End of eqalign
$$
\endproclaim

\beginproclaim{Corollary}{1}
The linear span of all normalized characters and supercharacters of
the minimal unitary discrete series with central charge $ 3 c_m$ is
invariant under  the action of $ SL_2 (\ZZ)$ defined by
$$
\pmatrix{ a & b \cr c & d} f (\tau, z) = e^{-{\pi i c z^2 \over c
   \tau + d} c_m} f
   \left(
      {a \tau + b \over c \tau + d}, {z \over c \tau + d}
   \right).
$$
\endproclaim

\beginexample{1}
$ m = 2\/$ (minimal possible).  Then $ A^+_2 = \{ (1, 0) \}\/$, $
A^-_2 = \left\{ \left( {1 \over 2}, {1 \over 2} \right) \right\}\/$,
and $ \ch^{(2, \epsilon)}_{j,k} = \sch^{(2, \epsilon)}_{j,k} =
1\/$.  Furthermore,
$$
\eqalign{
  \varphi^+_{1/2, 1/2} (\tau)
  & =
  \left(
    F^{(2)+}_{1/2, 1/2} (\tau, z)
    = \Psi^{(1/2)-} (\tau, z)
  \right)_{z = 0} = q^{1 \over 8} \Delta
  \left(
    q^{1 \over 2}
  \right)^2,
  \cr
  \varphi^-_{1/2, 1/2} (\tau)
  & =
  \left(
    F^{(2)-}_{1/2, 1/2} (\tau, z)
    = \Psi^{(1/2)+} (\tau, z)
  \right)_{z = 0} = q^{1 \over 8} \Delta
  \left(
    -q^{1 \over 2}
  \right)^2,
  \cr
  \varphi^-_{1, 0} (\tau)
  & =
  \left(
    F^{(2)-}_{1, 0} (\tau, z)
    = \Psi^{(0)+} (\tau, z)
  \right)_{z = 0} = {\textstyle {1 \over 2}} \squaretwo
  (-q)^2 . \cr
}% End of eqalign
$$
These three functions span a 3-dimensional
$ SL_2 (\ZZ)\/$-invariant space of modular forms of weight 1.
(Note that $ F^{(2)+}_{1,0} (\tau, z)\/$ has a pole at $ z = 0\/$.)
\endexample

\beginremarks{Remark}{1}
Similar results may be obtained for the affine denominator in the
general case.  As in [K5], this may be used in order to prove the
``strange formula'' (1.2).
\endremarks

\beginsection{7}{The case $ h^\vee = 0\/$}

The simplest non-trivial case when $ h^\vee = 0\/$ is $ \gg = A (1,
1)\/$.  It is more convenient to consider $ \gg = gl (2,2)\/$
instead, with simple roots $ \alpha_1\/$, $ \alpha_2\/$, $
\alpha_3\/$, such that $ (\alpha_1 | \alpha_1) = (\alpha_3 |
\alpha_3) = 0\/$, $ (\alpha_3 | \alpha_3) = 2\/$, $ - (\alpha_1 |
\alpha_2) = (\alpha_3 | \alpha_2) = 1\/$.  We conjecture that the
denominator identity still holds with an extra factor.  Namely,
letting $ x = e^{-\alpha_1}\/$, $ y = e^{-\alpha_2}\/$,
$ z = e^{-\alpha_3}\/$, introduce the following functions (fixed
under $ W \ltimes t_{\ZZ \alpha_2}\/$):
$$
f_n = x^{-n} z^{-n} (x^{2n - 1} - z^{2n - 1}) (x - z)
\sum_{j \in \ZZ_+} (-1)^j q^{(j+1) (j + 2n) / 2}.
$$

\beginproclaim{Conjecture}{1}
$\displaystyle
e^\rho \widehat\R = \left( 1 + \sum^\infty_{n=1} f_n \right) \sum_{n
\in \ZZ} t_{n \alpha_2} (e^\rho \R)\/$.
\endproclaim

If we divide both sides by $ \R\/$ and let $ x, y, z \rightarrow
1\/$ we obtain Jacobi's formula (0.4), {\it i.e.\/},
$$
\squaretwo_{8,n} = 16 \sum_{j|n} (-1)^{n+j} j^3.
$$

So far we have entirely left out the series of simple Lie
superalgebras $ \gg\/$ of type $ Q(m)\/$, which do not admit a
non-degenerate even invariant bilinear form (but admit an odd one),
hence it is natural to let $ h^\vee = 0\/$.  Its even part $
\gg_0\/$ is a simple Lie algebra of type $ A_m\/$ and the $
\gg_0\/$-module $ \gg_1\/$ is the adjoint $ A_m\/$-module.  Let $
\Delta\/$ (resp $ \Delta_+\/$) denote the set of roots (resp
positive roots) of $ A_m\/$, let $ \Pi = \{ \alpha_1, \ldots,
\alpha_m \}\/$ be the set
of simple roots, let $ \rho\/$ be the half of the sum of roots from
$ \Delta_+\/$, let $ W\/$ be the Weyl group, and let $ (.|.)\/$
be the $ W\/$-invariant bilinear form normalized by $ (\alpha_i |
\alpha_i ) = 2\/$.  Let $ s = \left[ {m + 1 \over 2} \right]\/$ and
consider the following (maximal orthogonal) set of $ s\/$ roots:  $
\gamma_1 = \alpha_1 + \ldots + \alpha_m\/$, $ \gamma_2 = \alpha_2 +
\ldots + \alpha_{m-1}, \ldots\/$.  The denominator identity for $ Q
(m)\/$ is given by the following formula:
$$
\R := \prod_{\alpha \in \Delta_+}
{1 - e^{-\alpha} \over 1 + e^{-\alpha}} =
\sum_{w \in W} \epsilon (w) w
\prod^s_{j=1} (1 + e^{-\gamma_j})^{-1}.
\leqno(7.1)
$$

The associated to $ Q(m)\/$ affine superalgebra $ \hat\gg\/$ is a
central extension (by $ \CC C\/$) of the Lie superalgebra $
\bigoplus_{j \in \ZZ} t^j \gg_{j \mod 2}\/$ (its Cartan matrix is
not symmetrizable).  We let $ q = -e^{-C}\/$ and define the affine
denominator as follows:
$$
\widehat\R = \R \prod^\infty_{n=1}
\left(
   \left(
      {1 - q^{2n} \over 1 - q^{2n-1}}
   \right)^{2s}
   \prod_{\alpha \in \Delta}
   {1 - q^{2n} e^\alpha \over 1 - q^{2n-1} e^\alpha}
\right).
$$
In order to state the affine denominator identity, define $
t_\alpha\/$ for $ \alpha \in \ZZ \Delta\/$ as follows
$$
t_\alpha (e^\beta) = q^{(\alpha | \beta)} e^\beta, \quad
\beta \in \ZZ \Delta, \quad
t_\alpha (q) = q.
$$

\beginproclaim{Conjecture}{2}
$\displaystyle
e^\rho \widehat\R =
\sum_{w \in W}\;
\sum_{\vec n \in \ZZ^s_+}\,
\sum_{\vec k \in \ZZ^s_+ \atop k_1 \geq \ldots \geq k_s}
\epsilon (w) q^{|\vec k|} wt_{n_1 \gamma_1 + \cdots + n_s \gamma_s}
\left(
   e^{\rho + \sum_i k_i \gamma_i}
\right)\/$.
\endproclaim

In particular, for $ m = 1\/$ Conjecture 7.2 reads
$$
\displaylines{
  \prod^\infty_{n=1}
    {(1 - q^{2n})^2 (1 - q^{2n-2} x) (1 - q^{2n} x^{-1})
    \over (1 - q^{2n-1})^2 (1 - q^{2n-1} x) (1 - q^{2n-1} x^{-1})} \cr
  \vbox{
    \halign{\hfill$\displaystyle#$ & $\displaystyle #$\hfill \cr
          = & \sum_{j,k \in \ZZ_+} q^{2jk + j + k} (x^{-j} - x^{j+1}) \cr
          = & \sum_{n \in \ZZ} {q^n \over 1 - q^{2n+1} x}, \cr
    } % End of halign
  } \cr% End of vbox
}% End of displaylines
$$
where $ x = e^{-\alpha_1}\/$.  This identity follows from (4.8) by
replacing $ q\/$ by $ q^2\/$, $ y\/$ by $ -qx\/$ and $ x\/$ by
$ -q\/$.  As we mentioned in Section 5, dividing both sides by $ 1 -
x\/$ and tending $ x\/$ to 1, we recover Legendre's formula $
\Delta (q)^4\/$.

For $ m = 2\/$, Conjecture 7.2 reads (here $ x = e^{- \alpha_1}\/$,
$ y = e^{- \alpha_2}\/$):
$$
\displaylines{
\displaystyle
\qquad\qquad\qquad\prod^\infty_{n=1}
{(1 - q^{2n})^2 (1 - q^{2n-2} x) (1 - q^{2n-2} y) (1 - q^{2n-2} xy)
\over (1 - q^{2n-1})^2 (1 - q^{2n-1} x) (1 - q^{2n-1} y) (1-q^{2n-1}
   xy) } \hfill \cr
\hfill \times {(1 - q^{2n} x^{-1}) (1 - q^{2n} y^{-1}) (1 - q^{2n}
   x^{-1} y^{-1}) \over (1 - q^{2n-1} x^{-1}) (1 - q^{2n-1} y^{-1})
   (1 - q^{2n-1} x^{-1} y^{-1})}\qquad\qquad\qquad \cr
=
\sum_{j \in \NN}\;
\sum_{k \in \NN {\rm odd}}
q^{jk-1} (xy)^{1-j} (1 - x^j) (1 - y^j) (1 - x^j y^j). \cr
}% End of displaylines
$$
Letting in this identity $ x = y = z\/$, we obtain after dividing by
$ (1 - z)^2 (1 - z^2)\/$:
$$
\matrix{
\displaystyle
\qquad\prod^\infty_{n=1}
{(1 - q^{2n})^2 (1 - q^{2n} z)^2 (1 - q^{2n} z^{-1})^2 (1 - q^{2n}
z^2) (1 - q^{2n} z^{-2}) \over
(1 - q^{2n-1})^2 (1 - q^{2n-1} z)^2 (1 - q^{2n-1} z^{-1})^2 (1 -
q^{2n - 1} z^2) (1 - q^{2n-1} z^{-2}) } \hfill \cr
\displaystyle
=
\sum_{j \in \NN}\;
\sum_{k \in \NN\, {\rm odd}} q^{jk - 1} z^{2 - 2j}
\left(
     {z^j - 1 \over z - 1}
\right)^2 {z^{2j} - 1 \over z^2 - 1}. \cr
}% end of matrix
$$
Letting $ z\/$ tend to 1, we obtain another Legendre's formula:
$$
\Delta (q)^8 =
\sum_{j \in \NN}\;
\sum_{k \in \NN\, {\rm odd}} j^3
q^{jk-1}, \qquad
\hbox{\it i.e.,} \quad
\Delta_{8, n} = \sum_{k | n + 1 \atop k {\rm odd}}
\left(
{n + 1 \over k}
\right)^3.
$$

Finally, dividing both sides of the identity for general $ m\/$ by
$ \R\/$ and evaluating at 0 using Weyl's character formula we obtain
the following two beautiful results for $ m = 2s - 1\/$ (resp.\ $
2s\/$):
$$
\Delta_{4s^2, n}
\left(
   \resp \Delta_{4s(s+1), n}
\right) =
\sum_{\vec r \in \ZZ^s_+}\;
\sum_{\vec k \in F_{\vec r, n}} \dim L
\left(
   k_1 \gamma_1 + \cdots + k_s \gamma_s, A_m
\right),
\leqno(7.2)
$$
where $ F_{\vec r, n} = \left\{ \vec k \in \ZZ^s_+ | k_1 \geq k_2
\geq \ldots \geq k_s,\; \sum^s_{i=1} ( k_i (2r_i + 1) + (m + 2 - 2i)
r_i ) = n \right\}\/$, \allowbreak
and $ L (\Lambda, A_m)\/$ denote the irreducible
(finite-dimensional) module over $ A_m\/$ with highest weight $
\Lambda\/$.

The simplest new formula corresponds to $ m = 3\/$:
$$
\Delta_{16, n} = {1 \over 12}
\sum_{{r_1, r_2 \in \ZZ_+ \atop k_1 \geq k_2 \in \ZZ_+} \atop k_1 (2
r_1 + 1) + k_2 (2 r_1 + 1) + 3 r_1 + r_2 = n}
(2 k_1 + 3) (2 k_2 + 1) (k_1 - k_2 + 1)^2 (k_1 + k_2 + 2)^2.
$$
Letting $ a = 2 k_1 + 3\/$, $ b = 2 k_2 + 1\/$, $ r = 2 r_1 + 1\/$,
$ s = 2 r_2 + 1\/$, we get the following beautiful form of this
formula:
$$
\Delta_{16, n} = {1 \over 3 \cdot 4^3}
\sum_{{a, b, r, s \in \NN {\rm odd} \atop a > b} \atop ar + bs = 2n
+ 4}
ab \left( a^2 - b^2 \right)^2.
$$
Next new formula corresponds to $ m = 4\/$:
$$
\Delta_{24, n} = {1 \over 72}
\sum_{{{{a, b \in \NN} \atop r, s \in \NN {\rm odd}} \atop a > b}
\atop ar + bs = n + 3}
a^3 b^3 \left( a^2 - b^2 \right)^2.
$$
In a similar way the general formula (7.2) can be rewritten as
follows:
$$
\eqalign{
\Delta_{4 s^2, n} & =
{4^{-s (s-1)} \over \prod^{2s - 1}_{j = 1} j!}
\sum_{{{{a_1, \ldots, a_s \in \NN {\rm odd} \atop r_1, \ldots, r_s
\in \NN {\rm odd}} \atop a_1 > \ldots > a_s} \atop a_1 r_1 + \cdots
+ a_s r_s = 2n + s^2}}
a_1 \ldots a_s \prod_{i < j} \left(a^2_i - a^2_j \right)^2 \cr
\Delta_{4 s (s + 1), n} & =
{2^s \over \prod^{2s}_{j = 1} j!}
\sum_{{{{a_1, \ldots, a_s \in \NN} \atop r_1, \ldots, r_s \in
\NN {\rm odd}} \atop a_1 > \ldots > a_s} \atop a_1 r_1 + \cdots +
a_s
r_s = n + {1 \over 2} s (s+1)}
(a_1 \ldots a_s)^3 \prod_{i < j} \left(a^2_i - a^2_j \right)^2. \cr
}% end of eqalign
$$

\beginsection{8}{Integrable highest weight modules over $
\hat\gg\/$}

Given a choice of $ \Delta_+ \subset \Delta\/$, we associate
to each $ \Lambda \in \hat\hh^\ast\/$ an {\it irreducible highest
weight module\/} $ \hat L (\Lambda)\/$ over $ \hat\gg\/$ defined
by the property that there exists a non-zero vector $ v_\Lambda\/$
such that:
$$
h (v_\Lambda) = \Lambda (h) v_\Lambda
\hbox{ for }
h \in \hat\hh, \quad
\nn_+ (v_\Lambda) = 0, \quad
(t^n \otimes \gg) v_\Lambda = 0
\hbox{ for }
n > 0.
$$
The number $ c = \Lambda (C)\/$ is called the {\it level\/} of $
\hat L (\Lambda)\/$.  Note that $ L (\Lambda) := U (\gg)
v_\Lambda\/$ is an irreducible highest weight module over $ \gg\/$.

\beginproclaim{Definition}{1}
A $ \hat\gg$-module $ \hat L (\Lambda)$ is called integrable if the
following two properties hold:
\itemitem{(i)} $ \dim L(\Lambda) < \infty$,
\itemitem{(ii)} $ t^n \otimes \gg_\alpha$ are locally nilpotent for
all
$ \alpha \in \Delta^\sharp_0$ and $ n \in \ZZ$.
\endproclaim

It is clear that if $ \hat L (\Lambda)\/$ is an integrable highest
weight $ \hat \gg\/$-module for some choice of $ \Delta_+\/$, then
it is an integrable highest weight module for any other choice of $
\Delta_+\/$; also, $ \hat L (\Lambda)\/$ is $ \gg\/$-locally finite.
The $ \hat\gg\/$-module $ \hat L (\Lambda)\/$ is integrable if and
only if $ \ch \hat L (\Lambda)\/$ is invariant with respect to $ W
\ltimes t_{M^\sharp}\/$.

In order to classify integrable highest weight $ \hat\gg\/$-modules
$\hat L (\Lambda)\/$, choose $ \Delta_+\/$ such that $ \Pi\/$
has a unique odd root ({\it cf.} Example 3.3).  Note that $
\alpha_0\/$ is even if and only if $ \gg\/$ is of the second type;
in this case define $ \tilde\theta \in \Delta_+ \subset
\hat\Delta_+\/$ as in Example~3.1c.  If $ \gg\/$ is of the
first kind, {\it i.e.,\/} of type $ A (m, n)\/$ or $ C (n)\/$, we
define $ \tilde\theta = \alpha_0 + \tilde\theta'\/$, where the
coefficients of $ \tilde\theta'\/$ in terms of $ \Pi\/$ are the
following:
$$
\matrix{
A (m, n),\ m \geq n\ &
\bigcirc - \bigcirc \cdots - \bigcirc - \bigotimes - \bigcirc_1 -
   \ldots - \bigcirc_1 \cr
C (m),\ m > 2 & \bigotimes^1 - \bigcirc-\bigcirc- \cdots -\bigcirc
\Leftarrow \bigcirc. \hfill \cr
}% End of matrix
$$
As before, we let $ k = ( \Lambda | \tilde\theta^\vee)\/$ and $ k_i
= ( \Lambda | \alpha^\vee_i )\/$ for all non-isotropic $
\alpha_i\/$.

\beginproclaim{Theorem}{1}
(a) For $ \gg$ of the second kind, the $ \hat\gg$-module $ \hat L
(\Lambda)$ is integrable if and only if $ \dim L (\Lambda) <
\infty$ (the corresponding conditions on $ k$ and the $ k_i$, $
i \neq 0\/$, are given in [K2, p.~84]) and $ k_0 \in \ZZ_+$.

(b) For $ \gg$ of type $ A (m, n)$, $ m \geq n$, the
$ \hat \gg$-module $ \hat L (\Lambda)$ is integrable if and only if
\itemitem{(i)} $ k \in \ZZ_+$ and $ k_i \in \ZZ_+$ for $ i \neq 0$
   and $ i \not= m + 1$;
\itemitem{(ii)} when $ k \leq n$, there exist $ r, s \in \ZZ_+$ such
   that: $ r + s = k$, $( \Lambda | \alpha_0 ) = r + k_{m + n + 1}
   + k_{m + n} + \cdots + k_{m + n - r + 2}\/$, $( \Lambda |
   \alpha_{m + 1}) = s + k_{m+2} + k_{m+3} + \cdots + k_{m + s +
   1}\/$.
\endproclaim

The $ \hat C (m)$-module $ \hat L (\Lambda)$ is integrable if and
only if
\itemitem{(i)} $ k \in \ZZ_+$ and $ k_i \in \ZZ_+$ for $ i \neq 0,
1$,
\itemitem{(ii)} if $ k = 0$, then $ ( \Lambda | \alpha_0) = (\Lambda
|
   \alpha_1) = 0$.

\beginproof
We use two facts:  (i) if $ \alpha$ is a simple even root, then the
root vector attached to $ -\alpha$ is locally nilpotent on $ \hat L
(\Lambda)$ if and only if $ ( \Lambda | \alpha^\vee ) \in \ZZ_+$;
(ii) any root from $ \widetilde\Delta$ can be made simple by applying
a sequence of odd simple reflections.
\endproof

One has the following expressions for the affine central charge $
c$:
$$
\vbox{\halign{$#$\hfil & \quad $#$\hfil \cr
A (m, n),\ m \geq n: & c = k_1 + \cdots + k_m + k. \cr
C (m): & c = k_2 + \cdots + k_m + k. \cr
B (m, n),\ D (m, n): & c = k_0 + k_1 + \ldots + k_{n-1} + k. \cr
D (2, 1; \alpha),\ F(4),\ G(3): & c = k_0 + k. \cr
}% End of halign
}% End of vbox
$$
It follows that the affine central charge $ c$ of an integrable $
\hat\gg$-module $ \hat L (\Lambda)$ is a non-negative integer and
that $ c = 0$ if and only if the $ \hat\gg$-module $ \hat L
(\Lambda)$ is 1-dimensional (which happens if and only if $ \Lambda
|_{\hh + \CC C} = 0$).

\beginexample{1}
$ \hat\gg = \hat A (1, 0)$.  We choose $ \Pi = \{ \alpha_1, \alpha_2
\}$ where both $ \alpha_1$ and $ \alpha_2$ are odd.  Then $
(\alpha_1 | \alpha_1) = (\alpha_2 | \alpha_2) = 0$, $ (\alpha_1 |
\alpha_2 ) = 1$.  Then $ \alpha_0 = C - \alpha_1 - \alpha_2$ is an
even root.  Let $ m_i = (\Lambda | \alpha_i)$, $ i = 0, 1, 2$.  One
has $ \dim L (\Lambda ) < \infty$ if and only if $ m_1 + m_2 \in
\NN$ or $ m_1 = m_2 = 0$, and $ \hat L (\Lambda)$ is integrable if,
in addition, $ m_0 \in \ZZ_+$.  The central charge $ c = m_0 + m_1
+ m_2$ ($ \in \ZZ_+$).  Assume that the following ``non-degeneracy''
condition holds:
$$
m_i \neq 0\ (i = 1 \hbox{ or } 2) \Rightarrow m_i
\hbox{ is not divisible by } c + 1.
$$
Then the same argument as in Section 2 gives the following character
formula:
$$
e^{\hat\rho} \widehat\R \ch \hat L (\Lambda) = \sum_{\alpha \in
M^\sharp} t_\alpha
\left(
e^{\hat\rho} \R \ch L (\Lambda)
\right).
\leqno(8.1)
$$
In particular, if $ m_1 = m_2 = 0$, we obtain for $ c \in \ZZ_+$:
$$
e^{\hat\rho} \widehat\R \ch \hat L (cd\,) =
\sum_{\alpha \in M^\sharp} t_\alpha
\left(
   e^{\hat\rho + cd} \R
\right).
\leqno(8.2)
$$
Letting $x = e^{- \alpha_1}\/$, $y = e^{- \alpha_2}\/$ we may
rewrite (8.2) as follows:
$$
e^{- cd} \widehat\R \ch \hat L (cd) =
\sum_{j, k \in \ZZ_+ \atop \min (j, k) \in (c+1) \ZZ}
        (-1)^{j+k} x^j y^k q^{jk \over c+1} -
\sum_{j, k \in \NN \atop \min (j, k) \in (c+1) \NN}
        (-1)^{j+k} x^{-j} y^{-k} q^{jk \over c+1} .
\leqno(8.3)
$$
\endexample

\vfill
\break

\beginproclaim{Definition}{2}
An integrable $ \hat\gg$-module $ \hat L (\Lambda)$ is called {\/\rm
tame} if its character is given by formula (8.1).
\endproclaim

Theorem 4.1 says that the 1-dimensional $ \hat\gg$-module is tame.

\medbreak%
\indent\indent{\bf References}

\frenchspacing

\item{[BL]}
   I. N. Bernstein, D.A. Leites, Character formul{ae} for
   irreducible representations of Lie superalgebras of series {\it
   gl\/} and {\it sl\/}, {\sl C.R. Acad. Bulg. Sci \bf33} (1980),
   \hbox{1049--51}.

\item{[D]}
   V.K. Dobrev, Characters of the unitarizable highest weight
   modules over the $ N = 2\/$ superconformal algebras, {\sl Phys.
   Lett. \bf 186B} (1987), \hbox{43--51}.

\item{[F]}
   N.J. Fine, {\sl Some basic hypergeometric series and applications},
   Math. Surveys {\bf27}, 1988, AMS, Providence.

\item{[H]}
   D. Hickerson, A proof of the mock theta conjectures, {\sl
   Invent. Math. \bf 94} (1988), \hbox{639--660}.

\item{[J]}
   C.G.J. Jacobi, Fundamenta nova theoriae functionum ellipticarum,
   {\sl Crelle J.} (1829), \hbox{55--239}.

\item{[JHKT1]}
   J. Van der Jeugt, J.W.B. Hughes, R.C. King and J. Thierry-Mieg,
   Character formul{ae} for irreducible modules of the Lie
   superalgebra $ sl(m/n)\/$, {\sl J. Math. Phys. \bf31} (1990),
   \hbox{2278--}.

\item{[JHKT2]}
   \leaders\hrule width2em\hskip2em, A character formula for singly
   atypical modules of the Lie superalgebra $ sl(m/n)\/$, {\sl
   Comm. Alg. \bf18} (1990), \hbox{3453--3480}.
{}.

\item{[K1]}
   V.G. Kac, Infinite-dimensional Lie algebras and Dedekind's $
   \eta\/$-function, {\sl Funct. Analis i ego Prilozh. \bf 8}
   (1974), No.~1, \hbox{77--78}.  English translation:  {\sl Funct.
   Anal. Appl. \bf8} (1974), \hbox{68--70}.

\item{[K2]}
   \leaders\hrule width2em\hskip2em, Lie superalgebras, {\sl
   Advances in Math. \bf26}, No.~1 (1977), \hbox{8--96}.

\item{[K3]}
   \leaders\hrule width2em\hskip2em, Characters of typical
   representations of classical Lie superalgebras, {\sl Comm. Alg.
   \bf5} (1977), \hbox{889--897}.

\item{[K4]}
   \leaders\hrule width2em\hskip2em, Representations of classical
   Lie superalgebras, {\sl Lect. Notes Math. \bf 676},
   Springer-Verlag (1978), \hbox{597--626}.

\item{[K5]}
   \leaders\hrule width2em\hskip2em, Infinite-dimensional algebras,
   Dedekind's $ \eta\/$-function, classical M\"obious function and
   the very strange formula, {\sl Advances in Math. \bf30} (1978),
   \hbox{85--136}.

\item{[K6]}
   \leaders\hrule width2em\hskip2em, Contravariant form for
   infinite-dimensional Lie algebras and superalgebras, in {\sl
   Lecture Notes in Physics \bf94}, Springer-Verlag (1979)
   \hbox{441--445}.

\item{[K7]}
   \leaders\hrule width2em\hskip2em, Laplace operators of
   infinite-dimensional Lie algebras and theta functions, {\sl
   Proc. Natl. Acad. Sci. USA \bf81} (1984), \hbox{645--647}.

\item{[K8]}
   \leaders\hrule width2em\hskip2em, Highest weight representations
   of conformal current algebras, Symposium on Topological and
   Geometrical methods in Field theory, Espoo, Finland, 1986.
   World. Sci., 1986, \hbox{3--16}.

\item{[K9]}
   \leaders\hrule width2em\hskip2em, Infinite-dimensional Lie
   algrebras, 3rd edition,  Cambridge University Press, 1990.

\item{[KL]}
   V.G. Kac, J. van de Leur, Super boson-fermion correspondence,
   {\sl Ann. d'Institute Fourier \bf37} (1987), \hbox{99--137}.

\item{[KP]}
   V.G. Kac, D.H. Peterson, Infinite-dimensional Lie algebras,
   theta functions and modular forms, {\sl Advances in Math. \bf53}
   (1984) \hbox{125--264}.

\item{[Ko]}
   B. Kostant, On Macdonald's $\eta$-function formula, the
   Laplacian and generalized exponents, {\sl Advances in
     Math. \bf20} (1976) \hbox{179--212}.

\item{[Mac]}
   I.G. Macdonald, Affine root systems and Dedekind's
   $ \eta\/$-function, {\sl Invent. Math. \bf15} (1972)
   \hbox{91--143}.

\item{[M]}
   Y. Matsuo, Character formula of $ c < 1\/$ unitary
   representation of $ N = 2\/$ superconformal algebra, {\sl Prog.
   Theor. Phys. \bf77} (1987), \hbox{793--797}.

\item{[P]}
   I. Penkov, Generic representation of classical Lie superalgebras
   and their localization, Univ. Cal. Riverside preprint (1992).

\item{[PS1]}
   I. Penkov, V. Serganova, Cohomology of $ G/P\/$ for classical
   Lie supergroups and characters of some atypical $ G\/$-modules,
   {\sl Ann. Inst. Fourier \bf39} (1989), \hbox{845--873}.

\item{[PS2]}
   \leaders\hrule width2em\hskip2em, Representations of classical
   Lie superalgebras of type I, {\sl Indag. Math \bf N.S.3(4)}
   (1992), \hbox{419--466}.

\item{[RY]}
   F. Ravanini, S.-K. Yang, Modular invariance in $ N = 2\/$
   superconformal field theories, {\sl Phys. Lett. \bf195B} (1987),
   \hbox{202--208}.

\item{[S]}
   V. Serganova, Kazhdan-Lusztig polynomials for the Lie
   superalgebras $ GL (m/n)\/$ Yale Univ. preprint (1992).

\item{[Se]}
   A. Sergeev, Tensor algebra of the defining representation as the
   module over Lie superalgebras $ gl (m,n)\/$ and $ Q(n)\/$, {\sl
   Math. USSR, Sbornik \bf51} (1985), \hbox{419--424}.

\item{[TM1]}
   J. Thierry-Mieg, Table des representations irreducibles des
   superalgebras de Lie, unpublished (1983).

\item{[TM2]}
   \leaders\hrule width2em\hskip2em, Irreducible representations of
   the basic classical Lie superalgebras $ SU (m/n)\/$, $ SU (n/n)
   /$ $U(1)\/$, $ OS p(m/2n)\/$, $ D(2/1,\alpha)\/$, $ G(3)\/$ and
   $ F(4)\/$, {\sl Lect. Notes in Physics \bf201}, eds. G. Denardo,
   G. Ghirardi and T. Weber, Springer, Berlin, \hbox{94--98}.

\item{[VJ1]}
   J. Van der Jeugt, Irreducible representations of the exceptional
   Lie superalgebras {\bf D}$ (2, 1:\alpha)\/$, {\sl J. Math. Phys.
   \bf26} (1985), \hbox{913--924}.

\item{[VJ2]}
   \leaders\hrule width2em\hskip2em, Character formul{ae} for the
   Lie superalgebra $ C (n)\/$, {\sl Comm. Alg.}

\bye